\documentclass[journal,singleside,final]{IEEEtran}

\usepackage{amsmath}
\usepackage{amsthm}
\usepackage{amsfonts}
\usepackage{graphicx}
\usepackage{enumitem}
\usepackage{color}
\usepackage[table]{xcolor}
\usepackage{float}
\usepackage{cite}
\usepackage{multirow}
\usepackage{graphics}
\usepackage[tight]{subfigure}
\usepackage{epsf}
\usepackage{epsfig}
\usepackage{amssymb}
\usepackage{comment}
\usepackage{epstopdf} 
\usepackage{bbold}
\usepackage[linesnumbered,ruled,vlined]{algorithm2e}
\SetKwInput{KwInput}{Input}                
\SetKwInput{KwOutput}{Output}              
\SetKw{Define}{Define}              
\SetKw{Continue}{continue}
\SetKw{Break}{break}
\SetKw{Return}{return}
\SetKwRepeat{Do}{do}{while}
\usepackage{algpseudocode}

\usepackage[skip=0pt,font=small]{caption}
\usepackage{color,soul}

\usepackage[hidelinks]{hyperref}
\definecolor{orange}{rgb}{1,0.5,0}
\definecolor{ao}{rgb}{0.0, 0.5, 0.0}

\usepackage[acronym,nonumberlist,style=super,toc]{glossaries}

\makenoidxglossaries
\newacronym{5g}{5G}{fifith generation}

\newacronym{ask}{ASK}{amplitude-shift keying}

\newacronym{awgn}{AWGN}{additive white Gaussian noise}
\newacronym{amc}{AMC}{adaptive modulation and coding}

\newacronym{ber}{BER}{bit error rate}
\newacronym{bler}{BLER}{block error rate}

\newacronym{bpsk}{BPSK}{binary phase-shift keying}
\newacronym{bs}{BS}{base station}
\newacronym{bsm}{BSM}{Basic Safety Messages}

\newacronym{csi}{CSI}{channel state information}
\newacronym{ch}{CH}{cluster head}
\newacronym{c-v2x}{C-V2X}{Cellular \gls{v2x}}
\newacronym{chs}{CHS}{\gls{ch} selection}
\newacronym{chs-pa}{CHS-PA}{\gls{chs} and \gls{pa}}

\newacronym{dl}{DL}{downlink}
\newacronym{dsrc}{DSRC}{Dedicated Short-Range Communications}
\newacronym{dnoma}{D-NOMA}{downlink-\gls{noma}}
\newacronym{dmnoma}{DM-NOMA}{downlink-M2M-\gls{noma}}

\newacronym{epa}{EPA}{equal power allocation}

\newacronym{fpa}{FPA}{fractional power allocation}
\newacronym{fd}{FD}{full-duplex}

\newacronym{cg}{CG}{cluster gateway}
\newacronym{gpa}{GPA}{greedy power allocation}

\newacronym{ht-urllc}{HT-URLLC}{high-throughput-\gls{urllc}}

\newacronym{iid}{i.i.d}{independent and identically distributed}
\newacronym{irs}{IRS}{intelligent reflecting surface}

\newacronym{jmld}{JMLD}{joint-multiuser maximum likelihood detector}

\newacronym{lut}{LUT}{look-up table}
\newacronym{los}{LOS}{line-of-sight}
\newacronym{llr}{LLR}{log-likelihood ratio}
\newacronym{lc}{LC}{lane cluster}
\newacronym{lte-v2x}{LTE-V2X}{Long-term Evolution-\gls{v2x}}

\newacronym{mld}{MLD}{maximum likelihood detector}
\newacronym{mrc}{MRC}{maximum ratio combining}
\newacronym{m2m}{M2M}{many-to-many}
\newacronym{m2mc}{M2MC}{\gls{m2m} communication}

\newacronym{md-chs}{MD-CHS}{minimum-distance \gls{ch} selection}
\newacronym{mimo}{MIMO}{multiple-input multiple-output}
\newacronym{mmWave}{mmWave}{millimeter-wave}

\newacronym{noma}{NOMA}{non-orthogonal multiple access}
\newacronym{nom}{NOM}{non-orthogonal multiplexing}
\newacronym{nlos}{NLOS}{non-\gls{los}}
\newacronym{ncg}{NCG}{non-\gls{cg}}
\newacronym{nch}{NCH}{non-\gls{ch}}

\newacronym{oma}{OMA}{orthogonal multiple access}
\newacronym{o-chs-pa}{O-CHS-PA}{optimal \gls{ch} selection and power allocation}
\newacronym{o-chs}{O-CHS}{optimal \gls{chs}}
\newacronym{opa}{OPA}{optimal power allocation}

\newacronym{pdf}{PDF}{probability density function}
\newacronym{pa}{PA}{power allocation}
\newacronym{pcb}{PCB}{power coefficient bound}
\newacronym{pam}{PAM}{pulse amplitude modulation}
\newacronym{p2p}{P2P}{peer-to-peer}

\newacronym{qos}{QoS}{quality of service}
\newacronym{qpsk}{QPSK}{quadrature phase-shift keying}
\newacronym{qam}{$M$-QAM}{$M$-ary quadrature amplitude modulation}
\newacronym{mqam}{QAM}{quadrature amplitude modulation}
\newacronym{mpam}{PAM}{pulse amplitude modulation}

\newacronym{rsu}{RSU}{roadside units}
\newacronym{rb}{RB}{resource block}

\newacronym{se}{SE}{spectral efficiency}
\newacronym{sc}{SC}{super cluster}
\newacronym{sch}{SCH}{\gls{sc} head}
\newacronym{scm}{SCM}{\gls{sc} member}
\newacronym{sic}{SIC}{successive interference cancellation}
\newacronym{sm}{SM}{spatial modulation}
\newacronym{snr}{SNR}{signal to noise ratio}
\newacronym{sinr}{SINR}{signal to interference plus noise ratio}

\newacronym{siso}{SISO}{single-input-single-output}
\newacronym{simo}{SIMO}{single-input-multiple-output}

\newacronym{scfp}{SCFP}{SC formation protocol}

\newacronym{ser}{SER}{symbol error rate}
\newacronym{s-chs-pa}{S-CHS-PA}{sub-optimal \gls{ch} selection and power allocation}

\newacronym{tdma}{TDMA}{time-division multiple access}

\newacronym{urllc}{URLLC}{ultra-reliable low-latency communication}
\newacronym{ul}{UL}{uplink}
\newacronym{cdf}{CDF}{cumulative  distribution  function}
\newacronym{unoma}{U-NOMA}{uplink-\gls{noma}}
\newacronym{umnoma}{UM-NOMA}{uplink-M2M-\gls{noma}}
\newacronym{udnoma}{UD-NOMA}{uplink-downlink-\gls{noma}}
\newacronym{udmnoma}{UDM-NOMA}{uplink-downlink-M2M-\gls{noma}}

\newacronym{vanet}{VANET}{vehicular ad hoc network}
\newacronym{v2x}{V2X}{vehicle-to-everything}

\renewenvironment{thebibliography}[1]{
  \begin{oldthebibliography}{#1}
    \setlength{\itemsep}{0.01em}
    \setlength{\parskip}{-0.12em}
}
{
  \end{oldthebibliography}
}
\begin{document}

\title{Novel Many-to-Many NOMA-based Communication Protocols for Vehicular Platoons}
\author{Mohammed~S.~Bahbahani,~ \IEEEmembership{Member,~IEEE,} Hamad~Yahya,~
\IEEEmembership{Member,~IEEE,} and Emad~Alsusa,~%
\IEEEmembership{Senior~Member,~IEEE,}
\thanks{Mohammed S. Bahbahani is with the Electronics Engineering Technology Department, College of Technological Studies, PAAET, Kuwait. (e-mail: \href{mailto:ms.bahbahani@paaet.edu.kw}{ms.bahbahani@paaet.edu.kw}).

Hamad Yahya is with the Department of Electrical Engineering, Khalifa University of Science and Technology, Abu Dhabi, 127788, UAE, (email: \href{mailto:hamad.myahya@ku.ac.ae}{hamad.myahya@ku.ac.ae}), and also with the Department of Electrical and Electronic Engineering, Imperial College London, London, SW7 2AZ, U.K. 

Emad Alsusa is with the Department of Electrical and Electronic Engineering, The University of Manchester, Manchester M13 9PL, U.K. (email: \href{mailto:e.alsusa@manchester.ac.uk}{e.alsusa@manchester.ac.uk})}}

\maketitle
\vspace{-10mm}
\begin{abstract}
\Gls{noma} is a promising technique for \acrlong{urllc} as it provides higher spectral efficiency and lower latency. In this work, we propose novel \gls{m2m} \gls{noma}-based schemes to exchange broadcast, multicast, and unicast messages between \glspl{ch} of vehicular platoons. Specifically, we design \gls{umnoma}, \gls{dmnoma} and joint \gls{udmnoma} schemes for \acrlong{p2p} \glspl{vanet}. We propose a unique clustering design for full-duplex communication that utilizes the high throughput \gls{mmWave} channels. Furthermore, we investigate jointly optimal \gls{chs} and \gls{pa} to maximize the network sum rate and devise a computationally efficient tailored-greedy algorithm that yields near-optimal performance. We also propose a super-cluster formation protocol to further limit the overhead of \gls{sic}. The results reveal that in most of the considered scenarios, the proposed \gls{udmnoma} scheme outperforms \gls{oma} in terms of sum rate by up to 50\% even when the \gls{sic} receiver errors reach 10\%. 
\end{abstract}

\glsresetall

\markboth{IEEE Journals,~Vol.~xx, No.~xx,
Jan~2024}{Bahbahani \MakeLowercase{\textit{et al.}}: Novel Many-to-Many NOMA-based Protocols for Future Vehicular Platoons} 

\IEEEpeerreviewmaketitle
\glsresetall
\section{Introduction}\label{sec:intro}

 \IEEEPARstart{T}{he} advent of \gls{v2x} wireless networks will revolutionize transportation safety and efficiency by enabling various vehicular applications, such as collision avoidance, traffic management, and eco-driving \cite{v2x_survey_2}. Vehicle platoons, whereby a line of cars drive closely with synchronized acceleration, have been shown to reduce carbon emissions and traffic congestion \cite{v2x_survey}, \cite{platoon_fuel}. 
 However, automated platoon driving requires \gls{urllc} as well as high data rates, which necessitate novel \gls{v2x} protocols and network designs \cite{vanet_survey1}. Specifically, platoons require low-latency communication to synchronize their speeds, accelerations, directions, and other critical parameters in real-time, as the inter-vehicle distance may be as small as 5 meters \cite{platoon_fuel}. Additionally, high volumes of data, including sensor information, control commands, and safety messages, must be reliably exchanged within the platoon and across neighboring platoons. 

Established \gls{v2x} technologies such as \gls{dsrc} \cite{dsrc_standard} and \gls{c-v2x} \cite{v2x_standards} have different topologies. While \gls{dsrc} operates in ad hoc mode to deliver basic safety messages with guaranteed channel access, \gls{c-v2x} leverages the cellular network infrastructure to enable \gls{v2x} applications. Besides, \gls{c-v2x} offers a side-link mode where vehicles can communicate directly without the \gls{bs}. Moradi-Pari \textit{et al.} \cite{dsrc_vs_clte} performed experimental performance comparisons between \gls{dsrc} and \gls{lte-v2x}. They concluded that \gls{lte-v2x} is superior for non-safety applications, whereas \gls{dsrc} is preferred for safety applications. While \gls{lte-v2x} suffers from limitations in latency due to handover, scheduling, and access grants, \gls{dsrc} struggles with low data rates, scalability, and reliability \cite{dsrc_vs_clte}. Thus, \gls{lte-v2x} and \gls{dsrc} cannot individually support safety and infotainment applications simultaneously, especially in highly dynamic and dense network scenarios.

To address the shortcomings of \gls{dsrc} and \gls{c-v2x}, researchers are exploring \gls{mmWave} communication \cite{mmWave_survey}, clustering \cite{cluster_survey}, \gls{m2mc}, and the application of \gls{noma} \cite{NOMA_vanet} for future \glspl{vanet}. Specifically, \gls{mmWave} communication enables multi-Gbps data rates due to the large available bandwidth in the $40$--$60$ GHz range, but at the expense of additional beam alignment and tracking overheads \cite{bahbahaniTDMA2022}. On the other hand, clustering improves scalability and reduces network congestion by aggregating common data at the \glspl{ch} \cite{aggregation_vanet,bahbahani_clustering}. The main literature that considered \gls{m2mc}, clustering protocols, and the application of \gls{noma} in vehicular communication is summarized in the following subsections. Interested readers can refer to \cite{bahbahaniTDMA2022} for further reading on vehicular communication with \gls{mmWave}.

\subsection{Directional Clustering Protocols for VANETs}
Recent clustering protocols for \glspl{vanet} have been highlighted in \cite{cluster_survey,cluster_survey2}. These protocols face many challenges, such as \gls{chs}, cluster formation, maintenance in a fast-changing topology, inter-cluster communication, etc. Zhang \textit{et al.} \cite{platoon_clustering} show that directional clustering can enhance cluster stability and extend link duration. Further, mmWave-based directional clustering is investigated in \cite{bahbahaniTDMA2022} and the references therein. In \cite{bahbahaniTDMA2022}, a directional clustering scheme is proposed whereby each vehicle of a platoon is linked to the cars in front and behind by directional \gls{mmWave} links. Packets are forwarded back and forth along the platoon cluster in a multi-hop fashion until received by the \gls{ch}, which then forwards aggregated data packets to nearby clusters via its sub-$6$ GHz omnidirectional antenna. Although the scheme supports high-throughput \gls{urllc} within the cluster, the lower data rates of the inter-cluster sub-$6$ GHz interface is a bottleneck, not to mention collisions and scheduling delays.

\subsection{NOMA for Vehicular Communication}
Power-domain \gls{noma} allows multiplexing multiple messages simultaneously in the same communication resource using distinct power levels \cite{Hamad2023-OJCOM}. Such an approach, however, is commonly associated with \gls{sic} at the receiver, which may present some practical challenges due to the added computational complexity of the receiver and the possibility of error propagation caused by \gls{csi} errors \cite{noma_decoding_order}. Notwithstanding, \gls{noma}'s gains outweigh its drawbacks as it provides improved spectral efficiency and user fairness, enables massive connectivity, and reduces latency and collisions, thereby facilitating \gls{urllc} for vehicular communications \cite{noma_myths,Jaiswal2021-TVT}. To address the practicality of \gls{sic} in large networks, hybrid TDMA-NOMA schemes have been proposed, in which a group of nodes (cluster) is assigned a time slot to transmit or receive their messages using \gls{noma} \cite{tdma-noma}.

Several works in the literature have comprehensively examined \gls{noma} in the uplink and downlink directions covering aspects such as \gls{pa} and modulation orders selection \cite{Hamad2022-TVT}, user pairing \cite{NOMA_pairing}, and cooperative relaying \cite{noma_relay}. For instance, Ding \textit{et al.} \cite{Ding2021-GLOBECOM} solve the resource allocation problem for a \gls{noma}-enabled platoon-based network by jointly considering the vehicle-to-vehicle and \gls{bs}-to-vehicle channels. Furthermore, Xiao \textit{et al.} \cite{noma_v2x_platoon} designed a \gls{noma}-enabled cluster formation and \gls{pa} algorithm to maximize the minimum data rate per \gls{ch}. Another \gls{pa} scheme is proposed in \cite{Guo2017-PIMRC} that targets the spectral efficiency and reliability of a heterogeneous \gls{v2x} network while considering imperfect channel estimation and in turn \gls{sic} error.

\subsection{Many-to-Many Communication} \label{sec:m2mc}
\gls{m2mc} allows multiple nodes to exchange their messages simultaneously in a decentralized fashion. In conventional cellular and cluster-based topologies, two users communicate by transmitting their packets to the \gls{bs} or \gls{ch} (many-to-one) which then forwards the packet to the intended receivers in the downlink (one-to-many). This scheme increases the latency while creating a bottleneck at the \gls{bs} or \gls{ch} \cite{m2m_vanet}. Additionally, failure of any of the two links corrupts the whole transmission, hence decreasing reliability. In contrast, a \gls{m2m} network can reduce latency while enhancing throughput and reliability \cite{m2m_clustering}. However, \gls{m2mc} entails several challenges such as coordination, synchronization, resource allocation, interference-management, and scalability, especially when \gls{oma} is adopted. Thus, combining \gls{noma} with \gls{m2mc}   may reap the gains from both technologies, while addressing \gls{m2mc} constraints.  

\begin{table}[]
    \centering
    \caption{List of abbreviations.}
\begin{tabular}{ |p{1.75cm}||p{6cm}|  }
 \hline
 \acrshort{awgn} & \Acrlong{awgn} \\
 \acrshort{bs} & \Acrlong{bs} \\
  \acrshort{c-v2x} & \Acrlong{c-v2x} \\
  \acrshort{ch} & \Acrlong{ch} \\
  \acrshort{chs} & \acrlong{chs} \\
  \acrshort{dnoma} & \Acrlong{dnoma} \\
  \acrshort{dsrc} & \Acrlong{dsrc} \\
  \acrshort{epa} & \Acrlong{epa} \\
  \acrshort{gpa} & \Acrlong{gpa} \\
  \acrshort{lc} & \Acrlong{lc} \\
  \acrshort{lte-v2x} & \Acrlong{lte-v2x} \\
  \acrshort{m2m} & \Acrlong{m2m} \\
  \acrshort{md-chs} & \Acrlong{md-chs} \\
  \acrshort{mmWave} & \Acrlong{mmWave} \\
  \acrshort{nch} & \Acrlong{nch} \\
  \acrshort{noma} & \Acrlong{noma} \\
  \acrshort{o-chs-pa} & \Acrlong{o-chs-pa} \\
  \acrshort{oma} & \Acrlong{oma} \\
  \acrshort{p2p} & \Acrlong{p2p} \\
  \acrshort{pa} & \Acrlong{pa} \\
  \acrshort{rb} & \Acrlong{rb} \\
  \acrshort{rsu} & \Acrlong{rsu} \\
  \acrshort{sc} & \Acrlong{sc} \\
  SCFP & Super-cluster Formation Protocol \\
  \acrshort{sch} & SC head \\
  \acrshort{s-chs-pa} & \Acrlong{s-chs-pa} \\
  \acrshort{sic} & \Acrlong{sic} \\
  \acrshort{sinr} & \Acrlong{sinr} \\
  \acrshort{snr} & \Acrlong{snr} \\
  \acrshort{tdma} & \Acrlong{tdma} \\
  \acrshort{udnoma} & \Acrlong{udnoma} \\
  \acrshort{unoma} & \Acrlong{unoma} \\
  \acrshort{urllc} & \Acrlong{urllc} \\
  \acrshort{v2x} & \Acrlong{v2x} \\
  \acrshort{vanet} & \Acrlong{vanet} \\
 \hline
\end{tabular}
    \label{tab:my_label}
\end{table}

\subsection{Motivations and Contributions}\label{sec:motivations}
Most research on \gls{noma} in the literature primarily focuses on cellular networks, considering either the uplink (many-to-one) \cite{tdma-noma}, \cite{baidas_d2d_noma} or downlink (one-to-many) \cite{Hamad2022-TVT} -- \cite{Xiao2019-IA} directions. However, the application of \gls{noma} in \gls{m2m} \glspl{vanet} has not been extensively investigated in the literature. Meanwhile, clustered vehicle platoons, such as the model proposed in \cite{bahbahani_clustering}, suffer from a bottleneck at the inter-cluster links. Hence, in this work, we propose novel schemes that integrate \gls{noma} with \gls{m2mc} to support URLLC and high throughput inter-cluster communication in platoon-based \glspl{vanet}.

The first scheme, \gls{dmnoma}, assigns a \gls{rb} to each \gls{lc} to transmit its messages to other \glspl{lc} multiplexed in the power domain. The second scheme, \gls{umnoma}, allocates a \gls{rb} to each \gls{lc} to receive messages from the other \glspl{lc} in an uplink \gls{noma} fashion. Unlike \gls{dmnoma} and \gls{umnoma}, which apply established uplink/downlink \gls{noma}, the third scheme, \gls{udmnoma}, is a novel application of \gls{noma} that integrates uplink and downlink \gls{noma} forming a two-layer \gls{noma} scheme. In one layer, each node combines its messages for the other nodes at specific power levels. In the second layer, the combined signals of all nodes are then superimposed again at different powers. As such, \gls{udmnoma} allows \glspl{lc} to exchange their messages over a single \gls{rb}.

Subsequently, we formulate a joint \gls{o-chs-pa} problem to maximize the network sum rate while maintaining a minimum data rate per \gls{lc} \cite{noma_downlink}. As the problem is complex and cannot be solved analytically, we obtain the optimal solution using MIDACO solver \cite{midaco}. Besides, we develop a sub-optimal solution based on a greedy algorithm that yields a near-optimal solution at lower complexity. We compare the proposed \gls{m2m}-\gls{noma} schemes against \gls{m2m} \gls{oma} using different power allocation and CH selection benchmark schemes. To the best of our knowledge, the proposed \gls{m2m} \gls{noma} schemes and the hybrid message-type transmission are unique. The contributions of this work can be summarized as follows:
\begin{itemize}
  \item Propose three schemes that utilize \gls{noma} in \gls{m2m} cluster-based \glspl{vanet}. While two schemes adapt existing hybrid OMA/NOMA directly to \gls{m2m} \glspl{vanet}, the third is a novel scheme that combines \gls{dmnoma} and \gls{umnoma} to achieve significantly higher data rates compared to OMA even with a moderate \gls{sic} error. These schemes leverage the advantages of \gls{m2mc} and \gls{noma} by minimizing latency while enhancing data rates, complementing existing C-V2X systems, and extending coverage in tunnels and suburban roads.   
  
  \item Propose a novel \gls{fd} inter-cluster communication model, in which the transmit and receive antennas are located on distinct vehicles within the cluster, utilizing high-speed mmWave links of the LC model in \cite{bahbahaniTDMA2022}. This approach reduces the complexity of \gls{fd} transceivers and self-interference cancellation due to the physical separation of the transmit and receive antennas.   

  \item Propose a hybrid \gls{m2m} NOMA-based data transmission scheme that allows unicast, multicast, and broadcast messages to be super-imposed in the same \gls{rb}, enabling various \gls{v2x} applications to coexist. It is particularly useful in scenarios where a \gls{lc} may broadcast a safety message while two \glspl{lc} exchange unicast messages \cite{unicast_multicast}. Extensive simulations show that for a given message type scenario, one of the proposed \gls{noma} schemes outperforms the benchmark.  
    
 \item Investigate the \gls{chs-pa} problem and design a low-complexity greedy algorithm that yields a near-optimal solution, offering a trade-off between performance and complexity.  

  \item Devise a distributed super-cluster formation protocol to limit the complexity of the \gls{chs-pa} problem solution and reduce the overhead of \gls{sic}. 

\end{itemize}

\begin{figure}[t]
    \centering
    {\includegraphics[width=3.8in,trim={0cm 0 0cm 0 },clip]{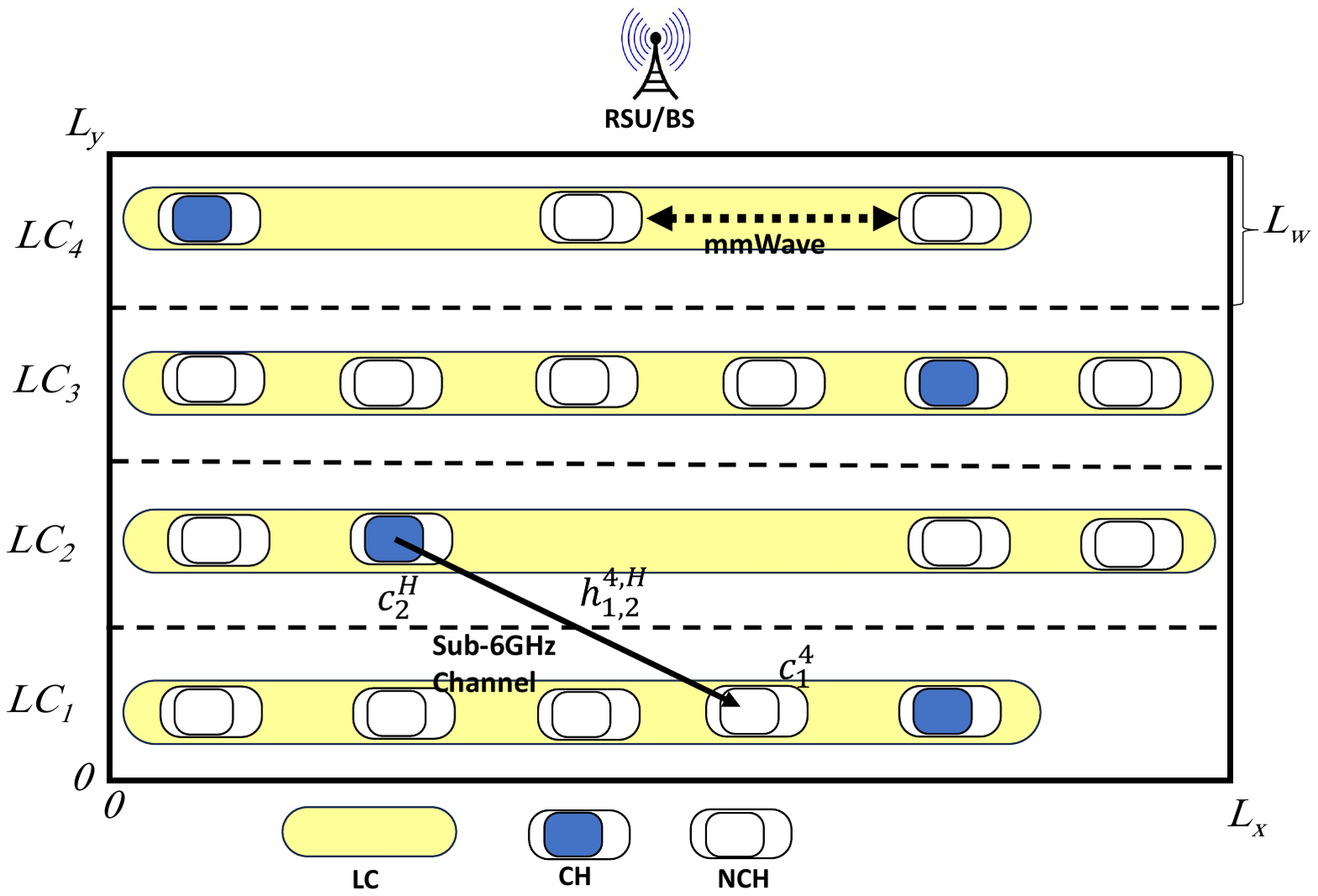}}
\caption{Network model block diagram.}
    \label{fig:systemModel}\vspace{-0.5mm}
\end{figure}
\subsection{Article Organization}
The remainder of the paper is organized as follows. The system model is described in Sec. \ref{sec:systemModel}. The signal analysis of the proposed \gls{m2m} \gls{noma} schemes is given in Sec. \ref{sec:proposed_noma}. In Sec. \ref{sec:Opt-CH-Select}, we formulate the joint \gls{o-chs-pa} problem, then present a low-complexity sub-optimal solution and analyze its complexity. Then, we describe the proposed super cluster formation protocol in Sec. \ref{sec:SCFP}. In Secs. \ref{sec:Results} and \ref{sec:Discussion}, we present and discuss the simulation results, respectively. Finally, concluding remarks and future work are highlighted in Sec. \ref{sec:Conclusions}.   
\section{System Model}\label{sec:systemModel}

\subsection{Network Model}

\begin{table}[]
    \centering
    \caption{Table of notations. }
\begin{tabular}{ |p{1.75cm}||p{6cm}|  }
 \hline
  $L_{x}$& Road length\\
  $L_{y}$& Road width\\
  $N_{L}$& Number of Lanes\\
  $N_{LC}$& Number of Lane Clusters\\
  $\mathcal{L}$& Set of LCs\\
  $\bold{c}_{i}$& Vector of LC $i$ members\\
  $N_{i}$& Size of LC $i$\\
  $c_{i}^{j}$& Vehicle $j$ or LC $i$\\
  $c_{i}^{H}$& CH of LC $i$\\
  $\bold{h}$& CH selection vector\\
  $B_{i}$& Broadcast/Multicast indicator variable of CH $i$\\
  $\bold{b}$& Vector of Broadcast/Multicast indicator variables $i$\\
  $M_{i,j}$& Binary indicator for a message from LC $j$ to LC $i$\\
  $\bold{M}$& Matrix of message indicator variables\\
  $h_{n,k}^{m}$& Channel gain of LC $k$'s CH to vehicle $m$ of LC $n$\\
  $\lambda$& Path loss exponent\\
  $P_{k}$& Transmit power of CH of LC $k$\\ 
  $P_{max}$& Total SC transmit power constraint\\
  $x_{n,k}$& Message signal of LC $k$ to LC $n$\\
  $w_{n,k}^{m}$& AWGN of message $x_{n,k}$ at vehicle $m$ of LC $n$\\
  $N_{o}$& Noise power spectral density\\
  $\gamma_{n,k}^{m}$& SNR of LC $k$'s message to vehicle $m$ of LC $n$\\
  $\gamma_{n,k}$& SNR of LC $k$'s message to LC $n$\\
  $W$& Channel bandwidth\\
  $R_{n,k}$& Data rate of LC $k$'s message to LC $n$ \\
  $R_{n,k}^{min}$& Minimum data rate constraint of message $x_{n,k}$ \\
  $f_{O}$& \gls{oma} bandwidth factor \\
  $f_{D}$& \gls{dmnoma} bandwidth factor \\
  $f_{U}$& \gls{umnoma} bandwidth factor \\
  $\alpha_{n,k}$& Fraction of power allocated to message $x_{n,k}$ \\
  $\boldsymbol{\alpha}$& Power allocation matrix \\
  $\zeta$ & Residual interference power factor \\
  $Q$ & Number of quantization level in GPA\\
  $N_{itr}^{max}$ & Maximum number of iterations in GPA\\

 \hline
\end{tabular}
    \label{tab:notations}
\end{table}

We consider a straight segment of a highway road of length $L_{x}$ and width $L_{y}$ that consists of $N_{L}$ lanes, in which $N_{LC}$ \glspl{lc} drive from left to right, as depicted in Fig. \ref{fig:systemModel}. A group of nearby \glspl{lc} within one-hop distance from each other forms a \gls{sc} by a protocol described in Sec. \ref{sec:SCFP}. Henceforth, we will consider a single \gls{sc} of size $N_{LC}$ denoted by the set $\mathcal{L}$ = \{$\bold{c}_{1}$,...,$\bold{c}_{N_{LC}}$\}. A \gls{lc} $\bold{c}_{i} \in \mathcal{L}$ represents a vector of $N_{i}$ vehicles denoted by $\bold{c}_{i}=(c^{1}_{i}$,...,$c^{N_{i}}_{i})$. Each \gls{lc} has a \gls{ch} (initially the front vehicle) denoted as $c^{H}_{i}\in{\bold{c}_{i}}$, which takes the role of transmitting the \gls{lc}s' messages to nearby \glspl{lc}. The remaining vehicles in the \glspl{lc}, called \gls{nch} vehicles, cooperatively receive the messages of other \glspl{lc} then forward them to the CH by \gls{mmWave} links as in \cite{bahbahaniTDMA2022}. We denote by $\bold{h}=(c^{H}_{1},...,c^{H}_{N_{LC}})$ the \gls{ch} selection vector. Each \gls{lc} may have:
\begin{enumerate}
    \item Unicast messages to a subset of the \glspl{lc} in the \gls{sc}: For example, a critical safety update or an entertainment content message is sent directly to a particular vehicle without overloading the network with unnecessary data traffic.  
    \item A multicast message to a subset of the \gls{sc}: the message can be a geocast safety message that targets \glspl{lc} in a specific location. For example, the \gls{lc} can send a geocast message to the \glspl{lc} behind when it is slowing down or send a geocast message to the \glspl{lc} ahead when accelerating. 
    \item A broadcast message to all \glspl{lc}: the message can be a traffic jam or car crash warning. 
    \end{enumerate}

    We define a matrix $\bold{M}$ of size $N_{LC}\times N_{LC}$, in which an element $M_{i,j}\in \{0,1\}$ is a binary variable that indicates if $\bold{c}_{j}$ has a message (unicast, multicast, or broadcast) to $\bold{c}_{i}$. Also, a vector $\bold{b}=(B_{1},...,B_{N_{LC}})$ such that $B_{i}\in\{0,1\}$ indicates if $\bold{c}_{i}$ has a broadcast/multicast message ($B_{i}=1$) or unicast messages ($B_{i}=0$) to the other \glspl{lc}. 

\subsection{Signal Model}\label{sec:signal-model}
We model the channel from a \gls{ch} $c_{k}^{H}$ to a \gls{nch} $c_{n}^{m}$ by a random variable $h_{n,k}^{m}=\hbar_{n,k}^{m}\times\hslash_{n,k}^{m}\times {d_{n,k}^{m}}^{-\lambda/2}$ that includes the small scale, large scale fading and knife-edge obstruction attenuation, where $\hbar\sim \mathcal{CN}(0,1)$, $d_{n,k}^{m}$ is the distance between the antennas of the two vehicles, and $\lambda$ is the path loss exponent. To obtain a realistic model of a highway vehicular network, we include $\hslash$ which is the attenuation caused by other vehicles obstructing the signal between the transmitter and receiver using the knife-edge model in \cite{knife-edge}.  

\begin{figure}[t]
    \centering
    {\includegraphics[width=3.5in,trim={0cm 0 0cm 0 },clip]{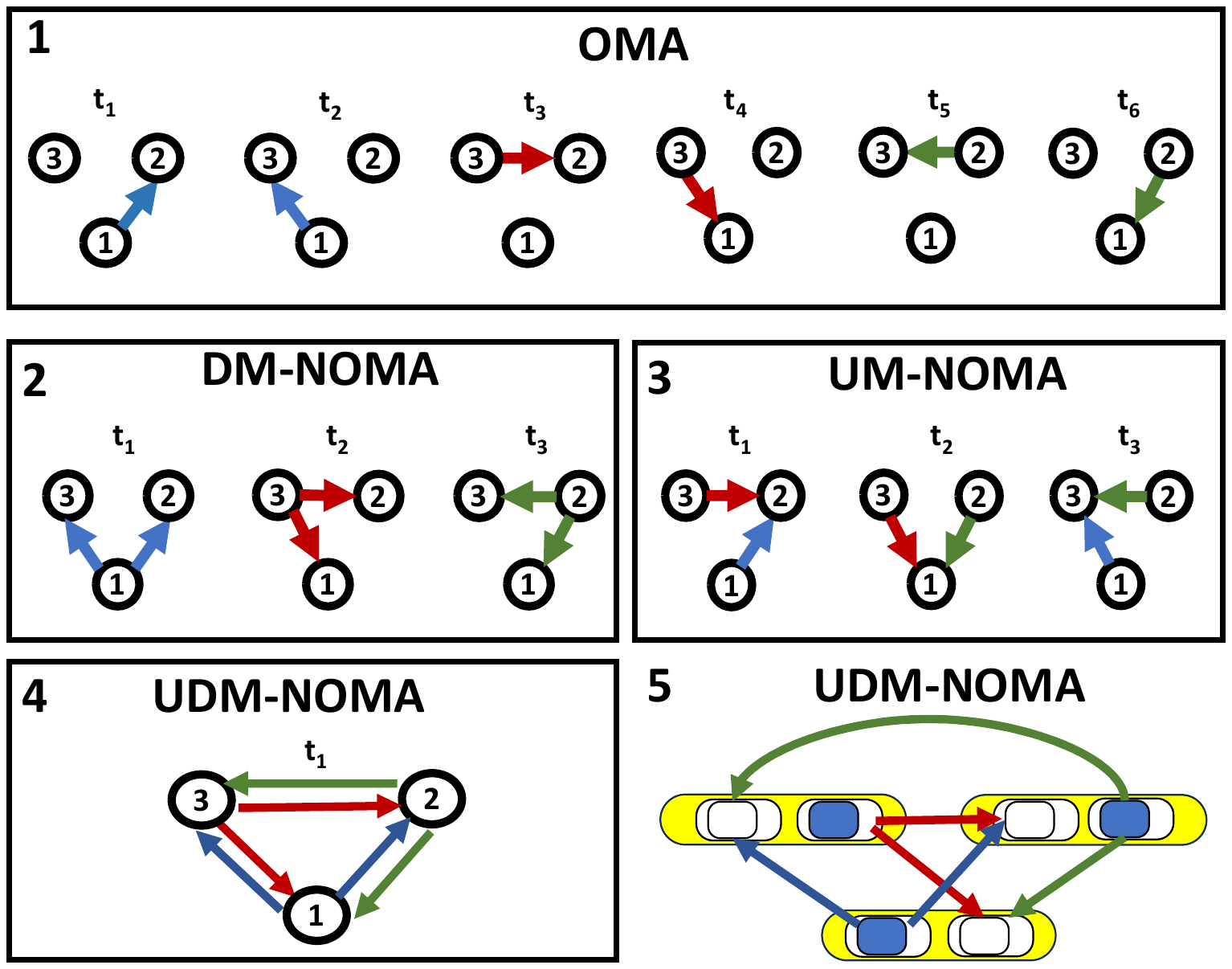}}
\caption{M2M MA schemes. The circles in 1-4 represent the LCs. Blue arrows indicate messages of LC 1, green arrows indicate messages of LC 2, and red arrows indicate messages of LC 3. }
    \label{fig:M2M_schemes}
\end{figure}

In the conventional \gls{oma} scheme, each \gls{ch} message is allocated an orthogonal \gls{rb}, such as a time slot. Hence, the message signal of \gls{ch} $c^{H}_{k}$ received by \gls{nch} $\mathit{c_{n}^{m}}$ can be given as
\begin{equation}
y_{n,k}^{m}=\sqrt{P_{k}}h_{n,k}^{m}M_{n,k}x_{n,k}+w_{n}^{m},\label{eq:yn_oma}
\end{equation}
where $P_{k}$ is the transmit power of $c_{k}^{H}$, $x_{n,k}$ is the message signal of $\bold{c}_{k}$ to $\bold{c}_{n}$, and $w_{n}^{m}\sim \mathcal{CN}(0,N_{0})$ is the \gls{awgn} at $\mathit{c_{n}^{m}}$ with noise power spectral density of $N_{0}$. Note that if $c_{k}^{H}$ is transmitting a broadcast message then $x_{1,k}=x_{2,k}=...=x_{N_{LC},k}=x_{k}$. Additionally, note that $\sum_{k}P_{k}=P_{\max}$ $\forall k$, where $P_{\max}$ is a total power constraint. In turn, the \gls{snr} at the receiver is given by
\begin{equation}
\gamma_{n,k}^{m}=\frac{M_{n,k}P_{k}|h^{m}_{n,k}|^{2}}{f_{O}N_{0}},\label{eq:SNR_oma}
\end{equation}
\begin{equation}
f_{O}=\frac{1}{\sum_{k=1}^{N_{LC}}(B_{k}+\overline{B_{k}}\sum_{n=1,n\neq k}^{N_{LC}}M_{n,k})},\label{eq:fo}
\end{equation}
where $f_{O}$ normalizes the bandwidth and noise power by dividing by the number of distinct transmitted messages. 
 $\overline{B_{k}}$ is the binary complement of $B_{k}$, that is, $\overline{B_{k}}=1-B_{k}$. Assuming the \gls{nch} signals are forwarded via perfect \gls{mmWave} links to the \gls{ch} by the protocol in \cite{bahbahaniTDMA2022}, the \gls{snr} of the message $x_{n,k}$, after performing selection combining among all received \gls{nch} signals, can be given as
\begin{equation}
\gamma_{n,k}=\max (\gamma_{n,k}^{1},...,\gamma^{N_{n}}_{n,k}).\label{eq:gamma_oma}
\end{equation}
Thus, the network sum rate becomes
\begin{equation}
R=f_{O}\sum_{n=1}^{N_{LC}}\sum_{k=1, k\neq n}^{N_{LC}} B_{k}\frac{R_{n,k}}{\sum_{j=1}^{N}M_{j,k}} +\overline{B_{k}}R_{n,k},\label{eq:sumR-oma}
\end{equation}
where $R_{n,k}=W\log_{2}(1+\gamma_{n,k})$ with $W$ being the channel bandwidth. Note that the data rate $R_{n,k}$ is normalized by $\sum_{j=1}^{N}M_{j,k}$ to achieve a fair comparison between unicast, multicast, and broadcast transmissions. Fig. \ref{fig:M2M_schemes}-1 illustrates the OMA scheme for a 3 LC network with each LC having a unicast message to the other two LCs requiring six time slots.

\section{Proposed M2M NOMA Schemes}\label{sec:proposed_noma}

In this section, we describe the three \gls{m2m} \gls{noma} schemes proposed in this article. 
\subsection{DM-NOMA Scheme}\label{sec:ol-noma-d-model}
In the \gls{dmnoma} scheme, each \gls{lc} is allocated an orthogonal \gls{rb} for transmitting a superposition of up to $N_{LC}-1$ messages to each other \gls{lc} using downlink NOMA. The signal of the \gls{ch} of $\bold{c}_{k}$ received at \gls{nch} $c_{n}^{m}$ is given as
\begin{equation}
y_{n,k}^{m}=\sqrt{P_{\max}}h_{n,k}^{m} \sum_{i=1, i\neq k}^{N_{LC}}M_{i,k}\sqrt{\alpha_{i,k}}x_{i,k}+w_{n}^{m}\label{eq:yn}
\end{equation}
where $0\leq\alpha_{i,k}\leq1$ is the fraction of $P_{\max}$ allocated to the message of $x_{i,k}$ such that $\sum_{k}\sum_{i}M_{i,k}\alpha_{i,k}=1$ $\forall i,k\in \{1,...,N_{LC}\}$\footnote{The self-interference signal transmitted by the \gls{ch} of LC $\bold{c}_{n}$ is assumed to have been perfectly canceled, as it can be conveyed to $c_{n}^{m}$ via high-speed \gls{mmWave} links.}. Note that for a broadcast/multicast message $\alpha_{1,k}=\alpha_{2,k}=...=\alpha_{N_{LC},k}=\alpha_{k}$. Assuming the availability of \gls{csi} and power coefficients, the receiver applies \gls{sic} to obtain the intended messages.
Let $\boldsymbol{\sigma}_{k}=(\sigma_{1,k},...,\sigma_{N_{LC},k})$ be a vector of received signal powers at $c_{n}^{m}$ from the CH of LC $\bold{c_{k}}$, where $\sigma_{i,k}=M_{i,k}P_{\max}|h^{m}_{n,k}|^{2}\alpha_{i,k}$ $\forall i\in{\{1,...,N_{LC}\}}$. Next, let $\hat{\boldsymbol{\sigma}_{k}}=(\hat{\sigma}_{1,k},...,\hat{\sigma}_{v-1,k},\hat{\sigma}_{v,k},\hat{\sigma}_{v+1,k},...,\hat{\sigma}_{N_{LC},k})$ be an ordered vector of $\boldsymbol{\sigma_{k}}$, such that $\hat{\sigma}_{v,k}=\sigma_{n,k}$, $\hat{\sigma}_{v,k}>\hat{\sigma}_{v+1,k}$ $\forall v\in\{1,..,N_{LC}-1\}$, where $\hat{\sigma}_{v,k} \in \boldsymbol{\sigma_{k}}$. Hence, the \gls{sinr} of the received message can be expressed as
\begin{equation}
\gamma_{n,k}^{m}=\frac{M_{n,k}P_{\max}|h^{m}_{n,k}|^{2}\alpha_{n,k}}{f_{D}\times(N_{0}+\underbrace{\sum_{j=v+1}^{N_{LC}}\hat{\sigma}_{j,k}}_\text{interference}+\underbrace{\sum_{j=1}^{v-1}\zeta\hat{\sigma}_{j,k}}_\text{SIC Error})},\label{eq:SNIR_n}
\end{equation}
\begin{equation}
f_{D}=\frac{1}{\sum_{k=1}^{N_{LC}}\bigvee_{j=1}^{N_{LC}}M_{j,k}},\label{eq:f_D}
\end{equation}
where $\bigvee$ is the logic OR operation between the $M_{j,k}$ binary variables, $\sum_{j=v+1}^{N_{LC}}\hat{\sigma}_{j,k}$ is the total interference power while decoding $x_{n,k}$, $\sum_{j=1}^{v-1}\zeta\hat{\sigma}_{j,k}$ is the residual power due to \gls{sic} error of the stronger signals, and $0\leq\zeta\leq1$ is a residual power factor, as in \cite{SIC_error}. The \gls{sinr} after performing selection combining as well as the sum rate expressions will be identical to (\ref{eq:gamma_oma}) and (\ref{eq:sumR-oma}). As shown in Fig. \ref{fig:M2M_schemes} 2), three LCs can exchange their messages over three time slots using \gls{dmnoma}.

\subsection{UM-NOMA Scheme}\label{sec:ol-noma-u-model}
In this scheme, the \glspl{lc} transmit their first messages together in a single \gls{rb} using uplink \gls{noma}. Hence, each \gls{nch} in this \gls{rb} will receive up to $N_{LC}-1$ messages, out of which one message is the intended one. In another \gls{rb}, each \gls{lc} transmits its next message, and so on. Thus, $N_{LC}$ orthogonal \glspl{rb} are used to transmit $N_{LC}(N_{LC}-1)$ messages, as depicted in Fig. \ref{fig:M2M_schemes} 3). Hence, the received signal at $c_{n}^{m}$ in RB $n$ becomes
\begin{equation}
y_{n}^{m}=\sqrt{P_{\max}}\sum_{k=1, k\neq n}^{N_{LC}}M_{n,k}h_{n,k}^{m}\sqrt{\alpha_{n,k}}x_{n,k}+w_{n}^{m}.\label{eq:yn-OL-U}
\end{equation}
 Here, we define $\boldsymbol{\sigma}_{n}=(\sigma_{n,1},...,\sigma_{n,N_{LC}})$ as a vector of all received signal powers at $c_{n}^{m}$ in \gls{rb} $n$, where $\sigma_{n,k}=M_{n,k}P_{\max}|h^{m}_{n,k}|^{2}\alpha_{n,k}$ $\forall k\in{\{1,...,N_{LC}\}}$. Similar to \gls{dmnoma} analysis, $\hat{\boldsymbol{\sigma}}_{n}=(\hat{\sigma}_{n,1},...,\hat{\sigma}_{n,v-1},\hat{\sigma}_{n,v},\hat{\sigma}_{n,v+1},...,\hat{\sigma}_{n,N_{LC}})$ is an ordered vector of $\boldsymbol{\sigma}_{n}$. Hence, the \gls{sinr} of the message of \gls{lc} $\bold{c}_{k}$ at \gls{nch} $c_{n}^{m}$ in \gls{rb} $n$ can be given respectively as
\begin{equation}
\gamma_{n,k}^{m}=\frac{M_{n,k}P_{\max}|h^{m}_{n,k}|^{2}\alpha_{n,k}}{f_{U}\times(N_{0}+\underbrace{\sum_{j=v+1}^{N_{LC}}\hat{\sigma}_{n,j}}_\text{interference}+\underbrace{\sum_{j=1}^{v-1}\zeta\hat{\sigma}_{n,j}}_\text{SIC error})},\label{eq:SNIR_n_ol_u}
\end{equation}
\begin{equation}
f_{U}=\frac{1}{\sum_{j=1}^{N_{LC}}\bigvee_{k=1}^{N_{LC}}M_{j,k}}.\label{eq:f_u}
\end{equation}
Again, the \gls{sinr} after selection combining and the sum rate are obtained using (\ref{eq:gamma_oma}) and (\ref{eq:sumR-oma}).

\subsection{UDM-NOMA Scheme}\label{sec:tl-noma-model}
Finally, we propose a novel combined uplink/downlink \gls{noma} scheme (\gls{udmnoma}), whereby all \glspl{ch} transmit their messages in the same \gls{rb} by applying uplink and downlink NOMA simultaneously. Hence, the received signal at \gls{nch} $\mathit{c_{n}^{m}}$ can be given as
\begin{equation}
y_{n}^{m}=\sqrt{P_{\max}}\sum_{k=1, k\neq n}^{N_{LC}}\!\!\!(h_{n,k}^{m}\sum_{i=1, i\neq k}^{N_{LC}}\!\!M_{i,k}\sqrt{\alpha_{i,k}}x_{i,k})+w_{n}^{m}.\label{eq:yn}
\end{equation}
The vector of received signal powers at $c_{n}^{m}$ becomes $\boldsymbol{\sigma}=(\sigma_{1,1},...,\sigma_{1,N_{LC}},\sigma_{2,1},...$ $,\sigma_{N_{LC},N_{LC}})$, where $\sigma_{i,k}=M_{i,k}P_{\max}|h^{m}_{n,k}|^{2}\alpha_{i,k}$ $\forall i,k\in{\{1,...,N_{LC}\}}$. Following the same analysis of the above schemes, $\hat{\boldsymbol{\sigma}}=(\hat{\sigma}_{1},...,\hat{\sigma}_{v-1}$ $,\hat{\sigma}_{v},\hat{\sigma}_{v+1},...,\hat{\sigma}_{N_{LC}\times N_{LC}})$ is an ordered vector of $\boldsymbol{\sigma}$, where $\hat{\sigma}_{v}={\sigma}_{n,k}$. Thus, the \gls{sinr} of a message received at $c_{n}^{m}$ of LC $\bold{c_{k}}$ becomes
\begin{equation}
\gamma_{n,k}^{m}=\frac{M_{n,k}P_{\max}|h^{m}_{n,k}|^{2}\alpha_{n,k}}{N_{0}+\underbrace{\sum_{j=v+1}^{N_{LC}}\hat{\sigma}_{j}}_\text{interference}+\underbrace{\sum_{j=1}^{v-1}\zeta\hat{\sigma}_{j}}_\text{SIC error}},\label{eq:SNIR_n}
\end{equation}
The \gls{sinr} after selection combining and sum rate analysis can be obtained as the other schemes above. Note that \gls{udmnoma} requires a single time slot, as shown in Fig. \ref{fig:M2M_schemes}-4 and Fig. \ref{fig:M2M_schemes}-5, and hence, a normalization factor is not required.  

\section{Cluster Head Selection and Power Allocation Problem Formulation}\label{sec:Opt-CH-Select}
The sum rate of the M2M NOMA schemes depends on the locations of CHs and the power assigned to each message. Hence, in this section, we tackle the joint \gls{o-chs-pa} problem for the proposed NOMA schemes. For a given set of $N_{LC}$ LCs, denoted by $\boldsymbol{\alpha}_{i}=(\alpha_{i,1},...,\alpha_{i,N_{LC}})$ the \gls{pa} vector of \gls{lc} $\bold{c}_i$, and $\boldsymbol{\alpha}=[\boldsymbol{\alpha}_{1}^{T},...,\boldsymbol{\alpha}_{N}^{T}]$ the \gls{pa} matrix, where $(\cdot)^{T}$ is the matrix transpose operation.
Hence, the problem is to find the optimal \gls{chs} vector $\bold{h}$ and the \gls{pa} matrix that maximize the sum rate while maintaining a minimum data rate $R_{n,k}^{\min}$ per message $x_{n,k}$. 
The optimization problem above is formulated as follows:
\begin{subequations}
\begin{equation}
\max_{\bold{h}, \boldsymbol{\alpha}} R,\label{eq:opt_1}
\end{equation}
s.t.
\begin{equation}
R_{n,k}\geq M_{n,k}R_{n,k}^{\min}, \forall n,k\in\{1,...,N_{LC}\},\label{eq:cons_1}
\end{equation}
\begin{equation}
\sum_{k=1}^{N_{LC}}\sum_{i=1}^{N_{LC}}M_{n,k}\alpha_{n,k}=1,\label{eq:cons_2}
\end{equation}
\begin{equation}
\alpha_{n,k}\geq 0, \forall n,k\in\{1,...,N_{LC}\},\label{eq:cons_4}
\end{equation}
\begin{equation}
    M_{n,k}\in\{0,1\}, \forall n,k\in\{1,...,N_{LC}\},\label{eq:cons_6}
\end{equation}
\begin{equation}
    B_{k}\in\{0,1\}, \forall k\in\{1,...,N_{LC}\},\label{eq:cons_7}
\end{equation}
\begin{equation}
    c^{H}_{i}\in \bold{c}_{i}, \forall i\in\{1,...,N_{LC}\},\label{eq:cons_8}
\end{equation}
\end{subequations}
Constraint (\ref{eq:cons_1}) requires that the data rate of each message is above the minimum data rate requirement. Constraint (\ref{eq:cons_2}) guarantees that the total transmit power of all \glspl{lc} is equal to $P_{\max}$. Constraint (\ref{eq:cons_4}) ensures that $\alpha_{n,k}$ is positive. Constraints (\ref{eq:cons_6}) and (\ref{eq:cons_7}) define the message availability, broadcast/multicast message, and \gls{chs} as binary variables, respectively. Finally, constraint (\ref{eq:cons_8}) indicates that, for each \gls{lc}, a \gls{ch} is selected from its members.  

\subsection{Sub-optimal Solution to the CHS-PA Problem}\label{sec:subOptSolution}
The \gls{chs-pa} problem is a non-convex mixed-integer nonlinear programming (MINLP) problem \cite{baidas_d2d_noma}, which is complex and cannot be solved in real-time to meet V2X latency requirements. Hence, we design a two-stage low-complexity sub-optimal solution, where the CHS and PA problems are solved independently as follows.

\subsubsection{Stage 1: FPA-based CHS}\label{sec:stage_1}
In the first stage, we evaluate the sum rate for every possible selection of CHs, assuming \gls{fpa}, whereby power is assigned proportional to channel gains \cite{fractional_PA}. The set of CHs that satisfies the minimum data rate constraints and maximizes the sum rate is then selected.

\subsubsection{Stage 2: Greedy Power Allocation Algorithm (GPA)}\label{sec:stage_1}

 After selecting the CHs in stage 1, PA is performed according to the proposed variation of the greedy algorithm listed in Algorithm \ref{alg:cap} below. First, each power variable in $\boldsymbol{\alpha}$ is discretized into $Q$ power levels, such that $\alpha_{n,k}\in\{0,\frac{1}{Q},\frac{2}{Q},...,1\}$. The power variables are initially set according to \gls{fpa}, rounding the real-valued powers to the nearest discrete power levels. PA is then performed by iteratively transferring a power level between pairs of power variables, as per Algorithm \ref{alg:OPSA}. Specifically, the sum rate resulting from decrementing a power variable (line 6) and incrementing another (line 7) is evaluated for every possible pair of variables. If $R^{min}$ constraints are not satisfied (line 8), the pair associated with the data rate $R_{n,k}$ furthest below $R^{\min}_{n,k}$ is selected (line 10-12). Otherwise, the pair that maximizes the sum rate without violating the constraints is selected (lines 16-18). Once a pair is found by Algorithm \ref{alg:OPSA}, the greedy algorithm (Algorithm \ref{alg:cap}) keeps transferring power between the optimal pair, one level at a time, until no gain in sum rate is obtained (lines 5-12). The greedy algorithm continues to search for another pair by Algorithm \ref{alg:OPSA} until no such pair exists or the maximum iterations $N_{itr}^{max}$ is reached.    
\subsection{Convergence of Proposed Greedy Algorithm} \label{sec:convergence}
We can show that the proposed greedy algorithm converges in a finite number of iterations as follows. Since the iterations continue only if a transfer between a pair or power variables strictly improves the sum rate, it is guaranteed that no power allocation can be visited twice. Given a finite number of power levels and, hence, a finite number of possible power allocations, the algorithm must eventually halt when either a local or a global optimal solution is reached.       
\subsection{Complexity of Proposed Sub-optimal Solution}\label{sec:complexity}
Obtaining the \gls{chs} through exhaustive search requires evaluating $\prod_{k=1}^{N_{LC}}N_{k}$ combinations, which grow exponentially in the number of \glspl{lc} and their sizes. Since \gls{fpa} has a complexity of order $O(1)$, the complexity of Stage 1 of the proposed sub-optimal \gls{chs-pa} solution becomes $O(N^{N_{LC}}_{k})$.
The complexity of the proposed greedy algorithm is analyzed as follows. In each iteration, $N^{4}_{LC}-2N_{LC}^{3}+N_{LC}$ combinations of pairs are evaluated. Although the number of iterations to convergence cannot be analytically obtained, its upper bound can be approximated to $\binom{Q+N_{LC}^{2}+N_{LC}-1}{N_{LC}^{2}+N_{LC}-1}$ which has high complexity. However, it will be shown in Sec. \ref{sec:Results} that the algorithm always converges in very few iterations when initialized with \gls{fpa}. Finally, note that the total complexity of the proposed sub-optimal solution is the sum of CHS and PA complexities. 

\begin{algorithm}[t]
\DontPrintSemicolon
  \KwInput{$\bold{G}$, $\mathcal{H}$, $\bold{M}$, $\bold{b}$, $\bold{R}^{\min}$, $\boldsymbol{\alpha}$, $Q$, $N_{itr}^{max}$}
  \KwOutput{$\boldsymbol{\alpha^{*}}$}
   $iter\leftarrow 0$\;
        \While{$iter<N_{itr}^{max}$}{
           $iter\leftarrow iter+1$\;
           $(Found,\boldsymbol{\alpha^{*}})\leftarrow OPSA(\frac{\boldsymbol{\alpha}}{Q},\bold{G},\bold{h},\bold{M},\bold{b},\bold{R}^{\min})$\;
           \uIf{$Found$}{
             $\alpha_{i^{*},j^{*}}\leftarrow \alpha_{i^{*},j^{*}}-1$,
             $\alpha_{k^{*},m^{*}}\leftarrow \alpha_{k^{*},m^{*}}+1$\;
             \While{$\alpha_{i^{*},j^{*}}>0$}{
             $\alpha_{i^{*},j^{*}}\leftarrow \alpha_{i^{*},j^{*}}-1$,
             $\alpha_{k^{*},m^{*}}\leftarrow \alpha_{k^{*},m^{*}}+1$\;
              $\bold{R^{new}}\leftarrow R(\frac{\boldsymbol{\alpha}}{Q},\bold{G},\bold{h},\bold{M},\bold{b},\bold{R}^{\min})$\;
              \uIf{$\min(\bold{R^{old}}-\bold{R^{min})}<0$ \& $\bold{R^{new}}>\bold{R^{old}}$}{
               $\bold{R^{old}} \leftarrow \bold{R^{new}}$
              }\uElse{
               $\alpha_{i^{*},j^{*}}\leftarrow \alpha_{i^{*},j^{*}}+1$,
               $\alpha_{k^{*},m^{*}}\leftarrow \alpha_{k^{*},m^{*}}-1$\;
                $Break$\;              
              }
             }
           }\uElse{
             $Break$\;
           }
        }

        $\boldsymbol{\alpha}^{*}=\boldsymbol{\alpha}$
        
\caption{Greedy power allocation algorithm.}
\label{alg:cap}
\end{algorithm}
\begin{algorithm}[t]
\caption{Optimal pair search algorithm (OPSA).}
\label{alg:OPSA}
     \KwInput{$\bold{G}$, $\bold{h}$, $\bold{M}$, $\bold{b}$, $\bold{R}^{\min}$, $\boldsymbol{\alpha}$, $Q$}
     \KwOutput{$\boldsymbol{\alpha^{*}}, Found$}
            $Found\leftarrow False$\;
          $\bold{R^{old}}\leftarrow R(\frac{\boldsymbol{\alpha}}{Q},\bold{G},\bold{h},\bold{M},\bold{b},\bold{R}^{\min})$\;
           $\bold{R_{diff}^{old}}\leftarrow \bold{R^{old}}-\bold{R^{min}}$\;

          \For{$\alpha_{i,j}$, $\alpha_{k,m} \in \boldsymbol{\alpha}$}{
              \uIf{$i\neq j$ $\&$ $k \neq m$ $\&$ $\sim (i=k$ $\&$ $j=m)$} {
                $\alpha_{i,j}\leftarrow \alpha_{i,j}-1$
                
                $\alpha_{k,m}\leftarrow \alpha_{k,m}+1$
                $\bold{R^{new}}\leftarrow R(\frac{\boldsymbol{\alpha}}{Q},\bold{G},\bold{h},\bold{M},\bold{b},\bold{R}^{\min})$\;
                \uIf{$\min(\bold{R^{old}_{diff}})<0$}{
                 $\bold{R_{diff}^{new}}\leftarrow \bold{R^{new}}-\bold{R^{min}}$\;
                  \uIf{$\min(\bold{R^{new}_{diff}})<$ $\min(\bold{R^{old}_{diff}})$}{
                  $\bold{R^{old}_{diff}}\leftarrow \bold{R^{new}_{diff}}$\;
                  $i^{*}\leftarrow i$, $j^{*}\leftarrow j$,  $k^{*}\leftarrow k$, $m^{*}\leftarrow m$
                  $Found \leftarrow True$\;   
                  } \uElse{
                 $\alpha_{i,j}\leftarrow \alpha_{i,j}+1$, 
                 $\alpha_{k,m}\leftarrow \alpha_{k,m}-1$
                 }
                }\uElse{
                 \uIf{$\bold{R^{new}}>\bold{R^{old}}$}{
                  $\bold{R^{old}}\leftarrow \bold{R^{new}}$\;
                  $i^{*}\leftarrow i$, $j^{*}\leftarrow j$,  $k^{*}\leftarrow k$, $m^{*}\leftarrow m$
                  $Found \leftarrow True$\;
                 }\uElse{
                 $\alpha_{i,j}\leftarrow \alpha_{i,j}+1$, 
                 $\alpha_{k,m}\leftarrow \alpha_{k,m}-1$
                 }
                
                }
               
              }
           }
\end{algorithm}

\begin{table*}[]
    \centering
    \caption{Summary of simulated scenarios.}
\begin{tabular}{lcccccccccccccccc}\hline
& \multicolumn{4}{||c||}{4LC-Unicast} & \multicolumn{4}{c||}{4LC-Broadcast} & \multicolumn{4}{c||}{4LC-Hybrid}
& \multicolumn{4}{c|}{3LC-Unicast} \\ \hline
& \multicolumn{1}{||c}{$1$} & \multicolumn{1}{|c}{$2$} & \multicolumn{1}{|c}{$3$} & \multicolumn{1}{|c||}{$4$} & \multicolumn{1}{c}{$1$} & \multicolumn{1}{|c}{$2$} & \multicolumn{1}{|c}{$3$} & \multicolumn{1}{|c||}{$4$} & \multicolumn{1}{c}{$1$} & \multicolumn{1}{|c}{$2$} & \multicolumn{1}{|c}{$3$} & \multicolumn{1}{|c||}{$4$} & \multicolumn{1}{c}{$1$}
& \multicolumn{1}{|c}{$2$} & \multicolumn{1}{|c}{$3$} & \multicolumn{1}{|c|}{$4$} \\ \hline\hline
\multicolumn{1}{c}{$1$} & \multicolumn{1}{||c|}{$\times$}  & \multicolumn{1}{c|}{$\checkmark $} & \multicolumn{1}{c|}{$\checkmark $} & \multicolumn{1}{c||}{$%
\checkmark $} & \multicolumn{4}{c||}{$\square $} & \multicolumn{1}{c|}{$\times $} & \multicolumn{1}{c|}{$\checkmark $}
& \multicolumn{1}{c|}{$\checkmark $} & \multicolumn{1}{c||}{$\checkmark $} & \multicolumn{1}{c|}{$\times $ }& \multicolumn{1}{c|}{$\checkmark $} & \multicolumn{1}{c|}{$\checkmark $}
& \multicolumn{1}{c|}{-} \\ \hline
\multicolumn{1}{c}{$2$} & \multicolumn{1}{||c|}{$\checkmark $} & \multicolumn{1}{c|}{$\times $} & \multicolumn{1}{c|}{$\checkmark $} & \multicolumn{1}{c||}{$\checkmark $} & \multicolumn{4}{c||}{$\square $} & \multicolumn{4}{c||}{$%
\square $} & \multicolumn{1}{c|}{$\checkmark $} & \multicolumn{1}{c|}{$\times $} & \multicolumn{1}{c|}{$\checkmark $} & \multicolumn{1}{c|}{-} \\ \hline
\multicolumn{1}{c}{$3$} & \multicolumn{1}{||c|}{$\checkmark $} & \multicolumn{1}{c|}{$\checkmark $} & \multicolumn{1}{c|}{$\times $} & \multicolumn{1}{c||}{$\checkmark $} & \multicolumn{4}{c||}{$\square $} & \multicolumn{1}{c|}{$\circ $} & \multicolumn{1}{c|}{$\times $}
& \multicolumn{1}{c|}{$\times $} & \multicolumn{1}{c||}{$\circ $} & \multicolumn{1}{c|}{$\checkmark $} & \multicolumn{1}{c|}{$\checkmark $} & \multicolumn{1}{c|}{$\times $} & \multicolumn{1}{c|}{-}
\\ \hline
\multicolumn{1}{c}{$4$} & \multicolumn{1}{||c|}{$\checkmark $} & \multicolumn{1}{c|}{$\checkmark $} & \multicolumn{1}{c|}{$\checkmark $} & \multicolumn{1}{c||}{$\times $} & \multicolumn{4}{c||}{$\square $} & \multicolumn{1}{c|}{$\times $} & \multicolumn{1}{c|}{$\times $} & \multicolumn{1}{c|}{$\times $} & \multicolumn{1}{c||}{$\times $} & \multicolumn{1}{c|}{-} & \multicolumn{1}{c|}{-} & \multicolumn{1}{c|}{-} & \multicolumn{1}{c|}{-} \\ \hline \\
\end{tabular}
\begin{tabular}{c}
      $\checkmark $: Unicast message, $\circ $: Multicast message, $\square $: Broadcast message, $\times $: No message
      \end{tabular}
\label{tab:summaryScen}\vspace{-7.5mm}
\end{table*}

\section{Proposed Super Cluster Formation and Overall Network Protocol}\label{sec:SCFP}
To manage the complexity of the CHS-PA problem and the \gls{sic} overhead in scenarios involving a large number of \glspl{lc}, and to support real-time performance, we propose a distributed \gls{scfp} by which a small group of LCs form and maintain a \gls{sc}, as follows.  

\subsection{Super Cluster Formation Protocol Description}\label{sec:SCFP_desc}
\begin{figure}[t]
    \centering
    {\includegraphics[width=3.8in,trim={0cm 0 0cm 0 },clip]{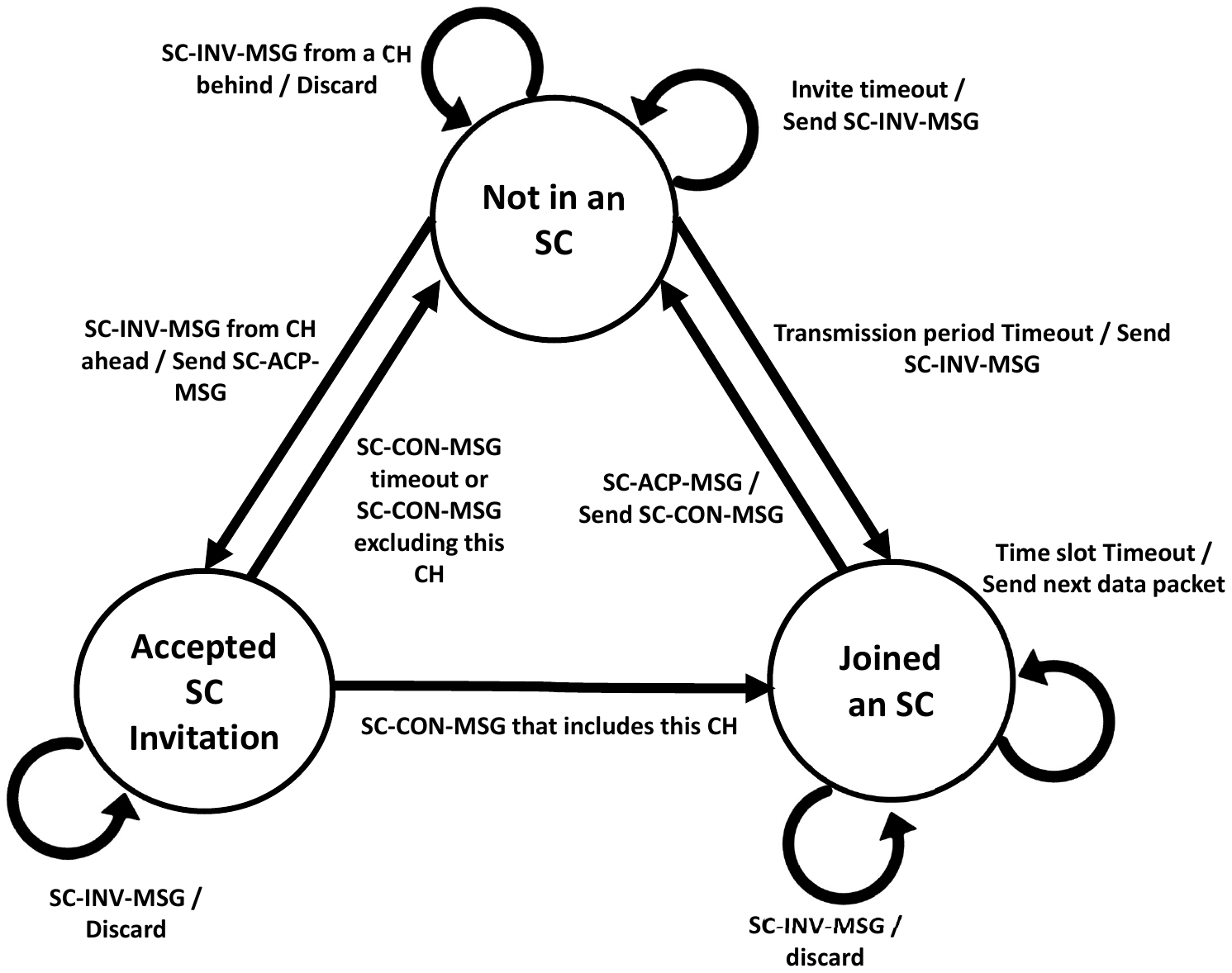}}
\caption{SCFP finite state machine.}
    \label{fig:SCFP_fsm}\vspace{-5mm}
\end{figure}

Before joining a \gls{sc}, a \gls{ch} periodically broadcasts an \gls{sc} invitation message SC-INV-MSG. A nearby \gls{ch}, not in any \gls{sc} accepts the invitation only if it is driving behind the inviting \gls{ch} to guarantee that all \glspl{ch} joining the \gls{sc} are within one hop distance\footnote{ It is assumed that \glspl{ch} have acquired their locations by GPS or other car positioning schemes \cite{positioning}. Also, note that the transmission range of sub-6GHz radios exceed 300 m in outdoor environments with NLOS links, allowing several platoons to become within one-hope range especially in highly dense scenarios.}. If a \gls{ch} accepts the invitation, it broadcasts an acceptance message SC-ACP-MSG that includes its ID and the locations of all its \glspl{nch}. Upon receiving the acceptance messages, the inviting \gls{ch} selects the nearest subset of the accepting \glspl{ch} below the \gls{sc} size limit. In turn, the inviting \gls{ch} becomes the \gls{sch}, and the accepting CHs as its \glspl{scm}. The \gls{sch} immediately broadcasts a confirmation message SC-CON-MSG that includes the CHS-PA vector. Note that \glspl{scm} are time-synchronized at the reception of the confirmation message. Notably, the \gls{sc} formation messages can be transmitted using the CSMA-CA protocol. The proposed SCFP is illustrated in Fig. \ref{fig:SCFP_fsm}.

\subsection{Resource Assignment in SCFP}\label{sec:SCFP_channel}
To maintain orthogonality between \gls{sc}s' transmissions, a \gls{ch} randomly selects a \gls{rb} from the available \glspl{rb} 
 and indicates it in the invitation message. Accepting \glspl{ch} specify the available \glspl{rb} in their acceptance messages after canceling the ones found in overheard invitations and confirmation messages. If the \gls{sch} discovers the \gls{rb} is reserved in acceptance messages, it selects another available \gls{rb} and then announces it in the confirmation message. Any \gls{rb} found reserved will be assumed available after a predetermined period. Hence, the protocol allows \glspl{rb} to be reused in distant \glspl{sc}.

\subsection{Overall Network Protocol}\label{sec:network-protocol-model}
Initially, vehicles are organized into \glspl{lc} following the protocol presented in\cite{bahbahani_clustering}. Subsequently, the established \glspl{lc} form \glspl{sc} according to the \gls{scfp} protocol explained in Sec. \ref{sec:SCFP_desc}. Each \gls{sch} computes then broadcasts the \gls{chs-pa} solution and SIC order to the \glspl{lc}, which then proceed to exchange data packets employing one of the \gls{noma} schemes detailed in Sec. \ref{sec:proposed_noma}. This exchange occurs over a period during which the channel gains and network topology are considered quasi-static \cite{bahbahaniTDMA2022}. Any alteration in the \gls{lc} topology triggers the \gls{lc} maintenance protocol described in \cite{bahbahani_clustering}, while modifications to the \gls{sc} structure re-invokes the \gls{scfp}.

\section{Numerical Results}\label{sec:Results}

In this section, we evaluate the performance of the proposed \gls{noma} schemes and sub-optimal \gls{chs-pa} using MATLAB. We consider a single \gls{sc} formed by the \gls{scfp} protocol described above. We set the road length and width to $L_{x}=60$ m and $L_{y}=12$ m to fit the topology in Fig. \ref{fig:systemModel}. All vehicles are assumed to be standard sedans with length, width, and height of $4.5$ m, $1.7$ m, $1.7$ m, respectively. The transmission frequency is set to $5.9$ GHz (Sub-6GHz DSRC frequency), and the path loss exponent is set to $\lambda=2.7$, a typical value for the sub-urban vehicular scenario. For all scenarios except the comparison in Subsec. \ref{snr_effect}, the SNR is set to 50 dB. The very high SNR value is used to evaluate the asymptotic performance of the proposed schemes in the absence of noise \cite{aymptotic_snr1}, \cite{aymptotic_snr2}. The minimum data rate constraint $R^{\min}$ for a message from a \gls{lc} in lane $i$ to a \gls{lc} in lane $j\neq i$ is set to $0.1\times|i-j|$ bps/Hz. As such, the further apart the LCs, the higher the guaranteed data rate, as higher relative speeds entail shorter-lived links\footnote{The value of 0.1 bps/Hz was chosen to assure the existence of a feasible solution to demonstrate the advantages of the proposed scheme. $R^{\min}$ can be linked to delay requirement, data size, etc. For example, a large message or a message that requires very low latency can be guaranteed a high data rate by increasing its $R^{\min}$ value compared to messages with less stringent latency requirements.}. Finally, the maximum number of iterations in GPA is set to $N_{itr}^{max}=100$.

In the performance evaluation, we consider the following topologies and transmission scenarios: 1) 4LC-Unicast, wherein a four-\gls{lc} network, each \gls{lc} sends a unicast message to every other \gls{lc}. 2) 4LC-Broadcast, wherein a four-\gls{lc} network, each \gls{lc} broadcasts a message to all other \glspl{lc}. 3) 4LC-Hybrid, wherein \gls{lc} 1 sends unicast messages to all other \glspl{lc}; \gls{lc} 2 broadcasts messages; \gls{lc} 3 sends multicast messages to \gls{lc} 1 and \gls{lc} 4, and \gls{lc} 4 does not send any messages. 4) 3LC-Unicast involves three \glspl{lc}, each sending a unicast message to the others. Table \ref{tab:summaryScen} summarizes the scenarios in a matrix, where each row is a transmitting \gls{ch} and each column is a receiving \gls{nch}. In all the simulated scenarios, we consider a single \gls{sc} with the vehicles positioned as in Fig. \ref{fig:systemModel} except for the three-\gls{lc} scenario in which the leftmost lane is omitted.  

\subsection{Performance of the Proposed NOMA Schemes}

\begin{figure}[t]
    \centering
    {\includegraphics[width=3.5in,trim={0cm 0 0cm 0 },clip]{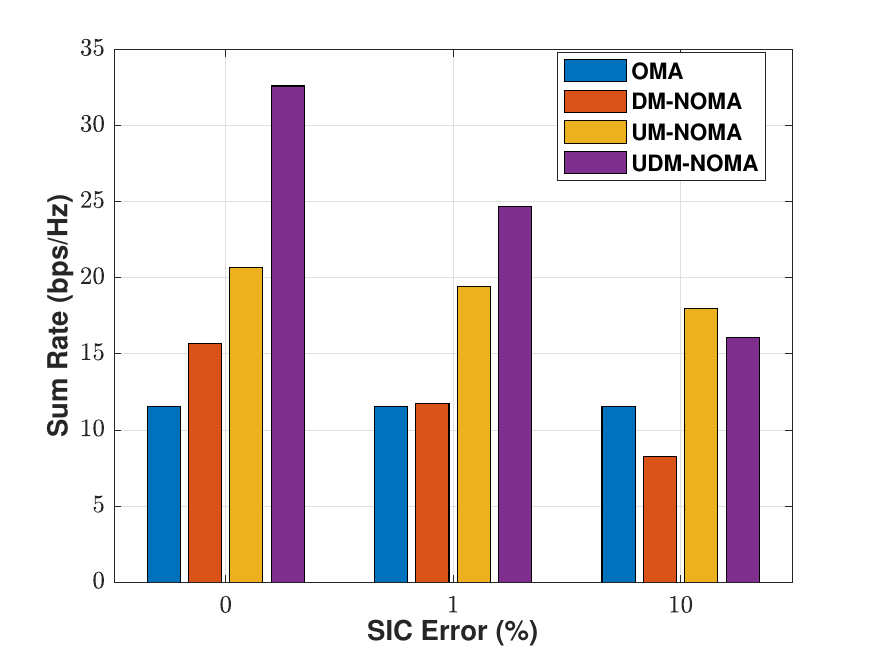}}
\caption{Sum rate vs. \gls{sic} error for 4LC-Unicast scenario.}
    \label{fig:sumrate_SIC_4LC_unicast}\vspace{-5mm}
\end{figure}

In this subsection, we compare the normalized sum rate ($W=1$ Hz) of the proposed \gls{m2m} \gls{noma} schemes against OMA in each of the topologies in Table \ref{tab:summaryScen} under optimal \gls{chs-pa} and with varying \gls{sic} errors. Fig. \ref{fig:sumrate_SIC_4LC_unicast} shows that, in the 4LC-Unicast scenario, the proposed \gls{noma} schemes significantly outperform \gls{oma} when \gls{sic} is perfect. Meanwhile, \gls{umnoma} always outperforms \gls{dmnoma}, since the signals in \gls{umnoma} arrive from different \glspl{lc}, thereby achieving higher diversity and more flexibility in power assignment compared to \gls{dmnoma}. Notably, \gls{udmnoma} achieves a superior sum rate due to more allocated bandwidth per message, achieving \gls{unoma} and \gls{dnoma} gains simultaneously. As the \gls{sic} error increases to 1\%, the performance of both \gls{umnoma} and \gls{udmnoma} deteriorates, albeit superior to \gls{oma}. A further increase in \gls{sic} error to 10\% adversely impacts \gls{udmnoma} due to the inclusion of additional interference terms, rendering \gls{umnoma} the most effective scheme \footnote{The 10\% \gls{sic} error is a very high value used to evaluate the asymptotic performance of the schemes. However, a 1-5\% \gls{sic} error is a practical value according to \cite{SIC_error} $[R_{1,1d}]$.}. 

\begin{figure}[t]
    \centering
    {\includegraphics[width=3.5in,trim={0cm 0 0cm 0 },clip]{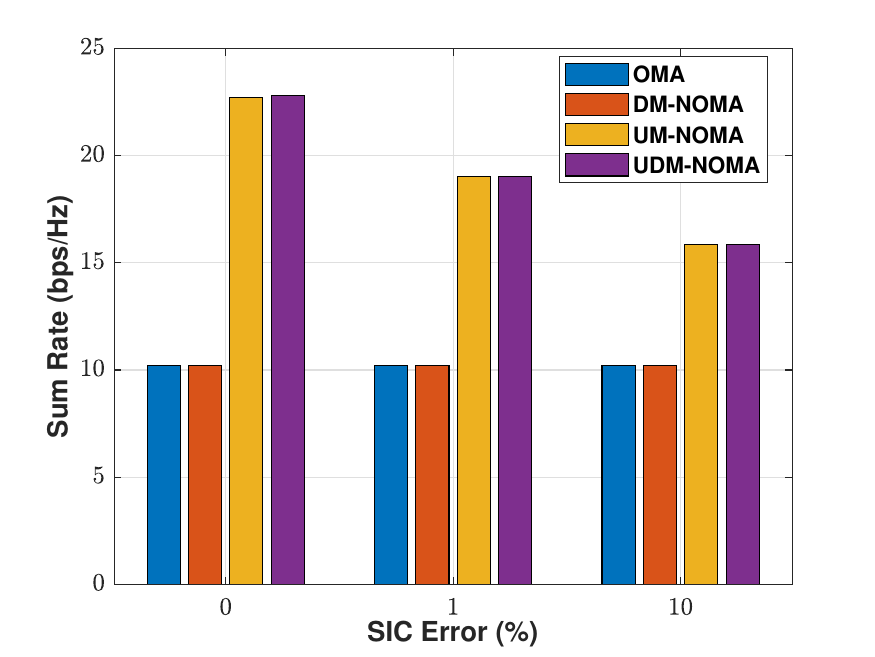}}
\caption{Sum rate vs. \gls{sic} error for 4LC-Broadcast scenario.}
    \label{fig:sumrate_SIC_4LC_broadcast}\vspace{-5mm}
\end{figure}

Subsequently, Fig. \ref{fig:sumrate_SIC_4LC_broadcast} illustrates the performance of the proposed schemes under the 4LC-Broadcast scenario. It is observed that \gls{dmnoma} attains an equivalent sum rate to \gls{oma}, remaining unaffected by \gls{sic} error. This is because each \gls{lc} transmits a singular broadcast message through a designated channel. Hence, \gls{noma} is not utilized with \gls{dmnoma} in the 4LC-Broadcast scenario as no interference exists. Conversely, \gls{umnoma} exhibits a considerable enhancement in performance, attributable to the multiplexing of multiple broadcast messages in the power domain. Notably, the performance of \gls{udmnoma} mirrors that of \gls{umnoma}, as only \gls{umnoma} is utilized in this scenario. 

\begin{figure}[t]
    \centering
    {\includegraphics[width=3.5in,trim={0cm 0 0cm 0 },clip]{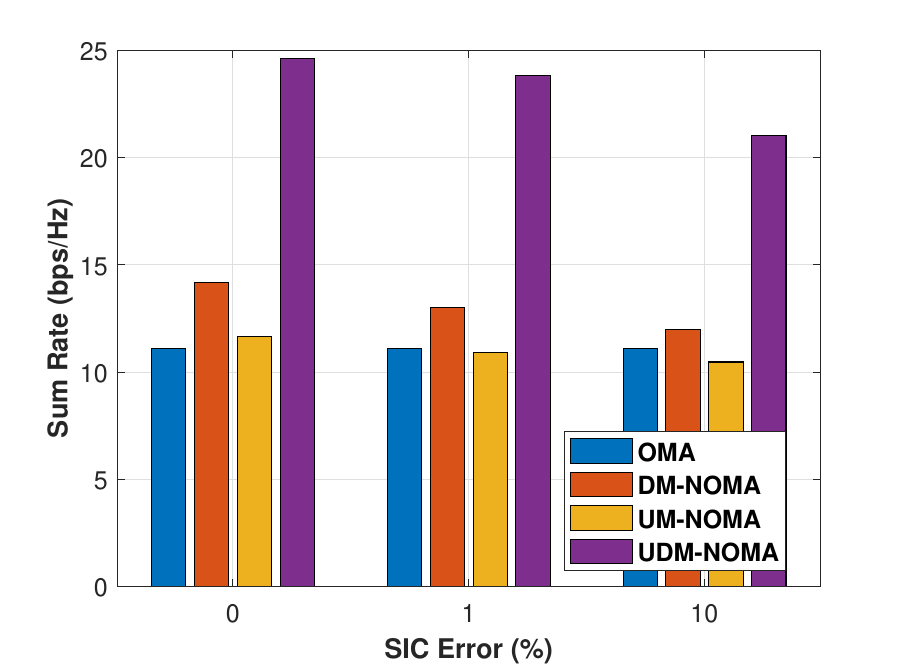}}
\caption{Sum rate vs. \gls{sic} error for 4LC-Hybrid scenario.}
    \label{fig:sumrate_SIC_4LC_hybrid}\vspace{-5mm}
\end{figure}

Fig. \ref{fig:sumrate_SIC_4LC_hybrid} depicts the performance of the 4LC-Hybrid scenario in Table \ref{tab:summaryScen} above. First, we note that the \gls{noma} schemes are less sensitive to \gls{sic} error compared to the unicast scenario. This is due to the fewer transmitted messages and, hence, interference terms in the 4LC-Hybrid scenario. Additionally, we observe that \gls{dmnoma} outperforms \gls{umnoma} in this scenario since three LCs are transmitting while four LCs are receiving. As such, \gls{umnoma} allocates four RBs to each receiving LCs while \gls{dmnoma} only allocates three RBs, multiplexing more messages in the power domain. With 0\% SIC error, \gls{umnoma} is marginally better than \gls{oma}, as it only utilizes one RB fewer than \gls{oma}, which requires five RBs. Meanwhile, \gls{udmnoma} surpasses all schemes in sum rate by fully utilizing the \gls{noma} gains.

\begin{figure}[t]
    \centering
    {\includegraphics[width=3.5in,trim={0cm 0 0cm 0 },clip]{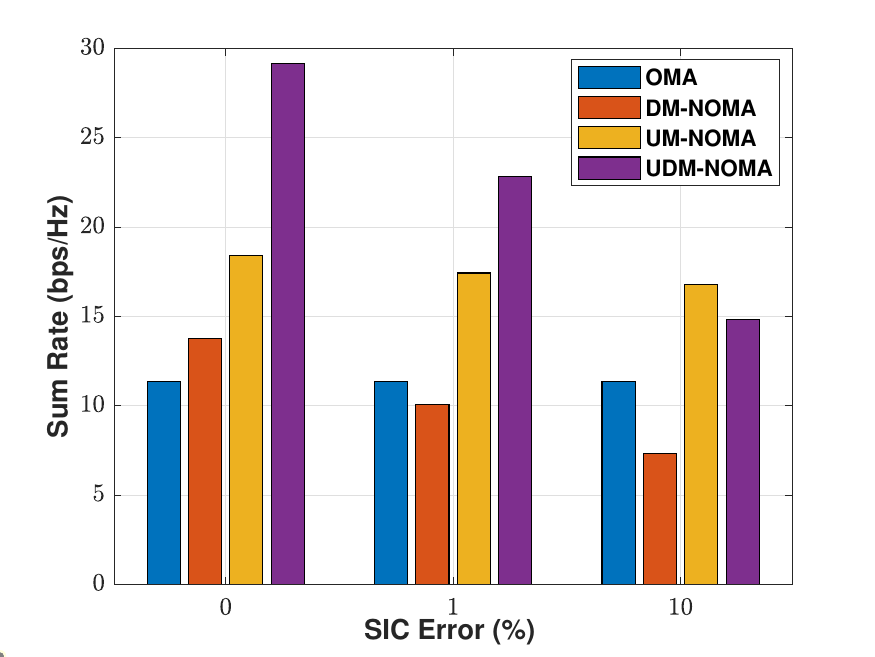}}
\caption{Sum rate vs. \gls{sic} error for 3LC-Unicast scenario.}
    \label{fig:sumrate_SIC_3LC_unicast}\vspace{-5mm}
\end{figure}

In the final simulated scenario depicted in Fig. \ref{fig:sumrate_SIC_3LC_unicast}, namely 3LC-Unicast, we observe the same trends as seen with the 4LC-Unicast scenario except that the gain in performance achieved by the three \gls{noma} schemes with respect to \gls{oma} is less significant as fewer messages are multiplexed in the power domain.

\subsection{Evaluation of Proposed Sub-optimal CHS-PA Scheme} \label{sec:eval_suboptimal_schemes}
\begin{figure}[t]
    \centering
    {\includegraphics[width=3.5in,trim={0cm 0 0cm 0 },clip]{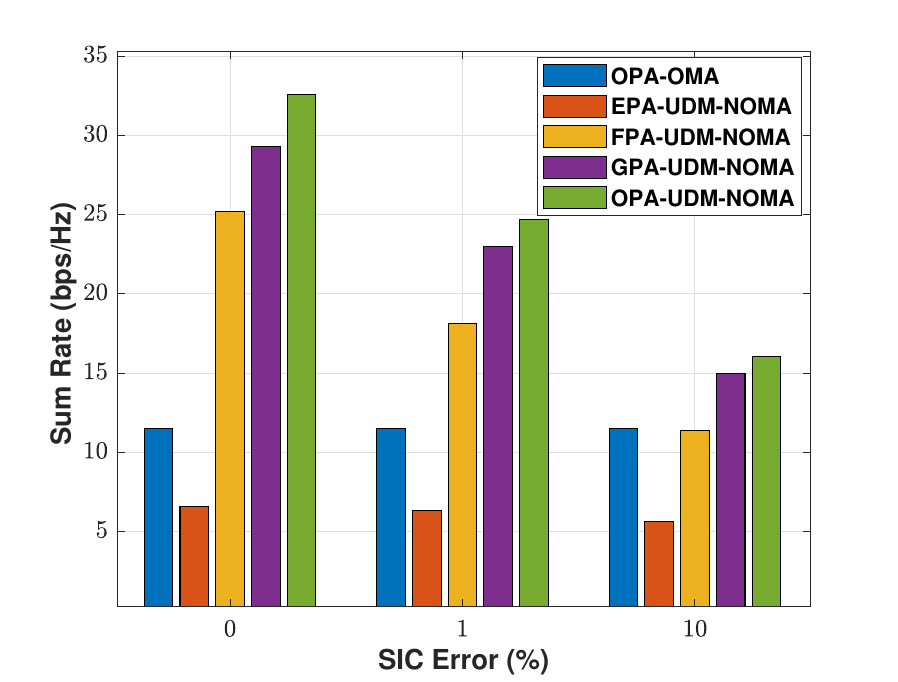}}
\caption{Sum rate of different PA schemes vs. \gls{sic} error (4LC-Unicast).}
    \label{fig:sumrate_PA_schemes}\vspace{-5mm}
\end{figure}

Fig. \ref{fig:sumrate_PA_schemes} depicts the sum rate achieved by the proposed sub-optimal \gls{gpa} solution compared to the following benchmark schemes: 1) \gls{opa} obtained by MIDACO Solver \cite{midaco}. 2) \gls{epa}, where power is divided equally between the transmitted messages, and 3) \gls{fpa}, where power is divided proportionally to channel gains, as in \cite{fractional_PA}. For all \gls{pa} schemes, CHS is solved by exhaustive search. For all schemes, except \gls{epa}, the proposed \gls{udmnoma} significantly outperforms \gls{oma} when the \gls{sic} error is low. The reason why \gls{udmnoma}'s sum rate deteriorates with \gls{epa} is the following. In \gls{dnoma}, messages transmitted at equal power experience the same channel gain, leading to high interference when implementing \gls{sic}. Since \gls{udmnoma} conducts both \gls{dnoma} and \gls{unoma} simultaneously, it encounters the same issue causing it to fall below the \gls{oma} sum rate.
It is also seen that the proposed sub-optimal solution is in close agreement with the \gls{opa}, but at a significantly reduced complexity. While surpassing \gls{fpa} by $14\%$, $27\%$, and $31\%$ , the proposed solution underperforms \gls{opa} by $9\%$, $8\%$, and $6\%$ for $0\%$, $1\%$, and $10\%$ \gls{sic} errors, respectively. This is due to the intricate nature of the \gls{udmnoma} problem surpassing the capabilities of the greedy algorithm, which may fall into local optima.

\subsection{Effect of Optimal CH Selection}\label{sec:effect_ochs}
\begin{figure}[t]
    \centering
    {\includegraphics[width=3.5in,trim={0cm 0 0cm 0 },clip]{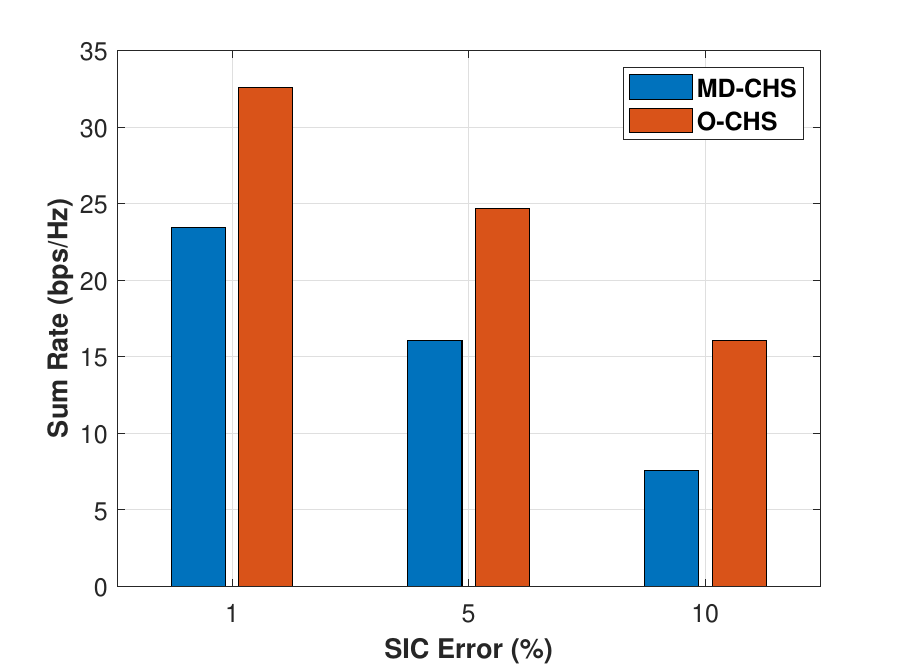}}
\caption{Sum rate of optimal and sub-optimal \gls{chs} vs. \gls{sic} error for \gls{udmnoma} (4LC-Unicast).}
    \label{fig:sumrate_CH_selection}\vspace{-5mm}
\end{figure}

Here, we investigate the effect of CH selection on the sum rate while power is optimally assigned. Specifically, we compare \gls{o-chs}, obtained with exhaustive search, against a non-optimal CHS, namely the \gls{md-chs}, in which the CHs closest to each other are selected, assuming the 4LC-Unicast scenario and \gls{opa}. In the simulated scenario, the \gls{md-chs} is $\{c_{5}^{1},c_{5}^{2},c_{5}^{3},c_{5}^{4}\}$, whereas the optimal \glspl{ch} are $\{c_{2}^{1},c_{1}^{2},c_{6}^{3},c_{5}^{4}\}$. With perfect \gls{sic}, Fig. \ref{fig:sumrate_CH_selection} shows that \gls{o-chs} outperforms \gls{md-chs} by $40\%$. This is due to the more distinct channel gains between the optimally selected \glspl{ch} and the receiving \glspl{nch}. As \gls{sic} error increases to $10\%$, \gls{o-chs} becomes more critical as the gain rises to nearly $100\%$ over \gls{md-chs}.       

\subsection{Effect of Number of Quantization Levels}

\begin{figure}[t]
    \centering
    {\includegraphics[width=3.5in,trim={0cm 0 0cm 0 },clip]{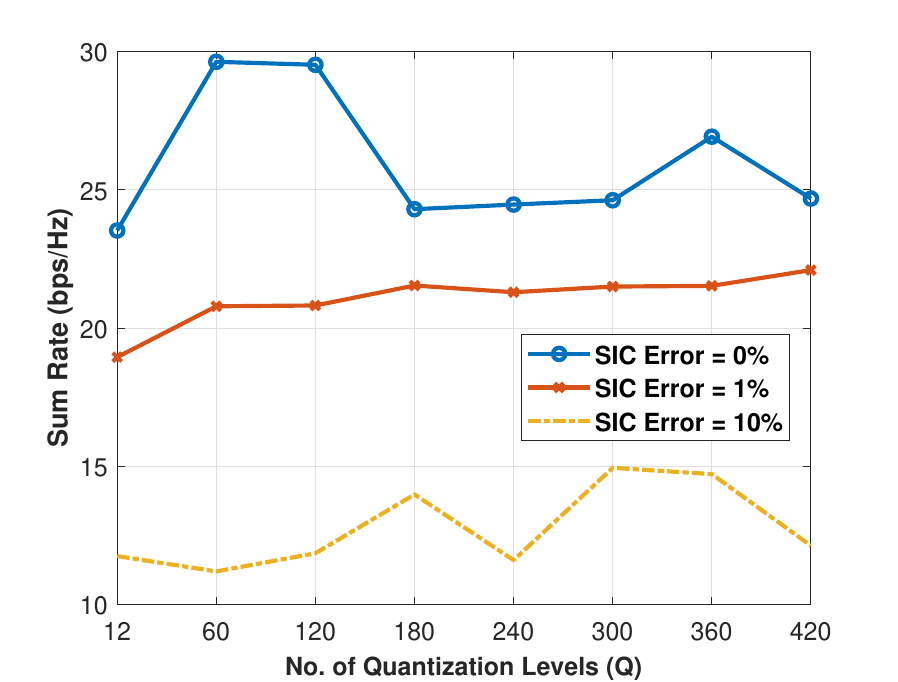}}
\caption{Sum rate vs. quantization levels for \gls{udmnoma} (4LC-Unicast).}
    \label{fig:sumrate_vs_Q_TL_4LC_unicast}
\end{figure}

In Fig. \ref{fig:sumrate_vs_Q_TL_4LC_unicast}, we explore the effect of changing the number of quantization levels $Q$ on the performance of the greedy algorithm. Starting with $Q=N_{LC}(N_{LC}-1)$ ($Q=12$ for $N_{LC}=4$), $Q$ is increased in multiples of 12 to a very high value ($Q=420$). The results show that the sum rate occasionally drops as $Q$ increases. This is because the amount of power increase per step of the iterative algorithm affects the \gls{sic} order. Specifically, increasing the power by a higher value (lower $Q$) causes the \gls{sinr} of the message to be decoded earlier in the \gls{sic} process hence suffering from more interference terms of the other transmitted messages and vice versa. We observe that a near-optimal value of $Q$ is always found for $Q<180$ for all \gls{sic} errors. Beyond $Q=180$, only a marginal performance enhancement is obtained at a higher complexity cost.  

\begin{figure}[t]
    \centering
    {\includegraphics[width=3.5in,trim={0cm 0 0cm 0 },clip]{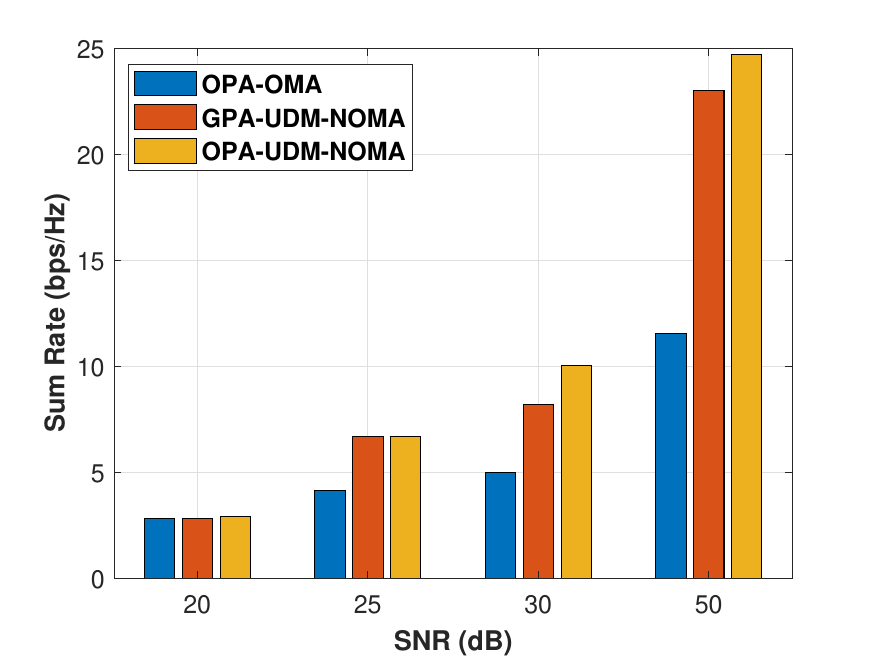}}
\caption{Sum rate vs. transmit \gls{snr} for \gls{udmnoma} with \gls{sic} error = $1\%$ (4LC-Unicast).}
    \label{fig:sumrate_vs_SNR_4LC_unicast}
\end{figure}

\subsection{Low-SNR vs. High-SNR} \label{snr_effect}

Finally, we compare the performance of \gls{udmnoma} with respect to \gls{oma} in high, moderate, and low \gls{snr} regions with $1\%$ SIC error. As depicted in Fig. \ref{fig:sumrate_vs_SNR_4LC_unicast}, lowering the transmit \gls{snr} reduces the \gls{udmnoma} to \gls{oma} gain, which drops from $115\%$ at \gls{snr} of $50$ dB to $63\%$ and $0\%$ at $25$ dB and $20$ dB, respectively, as generally the case in NOMA \cite{Hamad2021-IoT}. The results also show that the gap between \gls{opa} and \gls{gpa} for \gls{udmnoma} increases as transmit \gls{snr} exceeds $30$ dB.

\subsection{Complexity of Proposed Sub-optimal CHS-PA Solution}\label{sec:complexity_results}
\begin{figure}[t]
    \centering
    {\includegraphics[width=3.5in,trim={0cm 0 0cm 0 },clip]{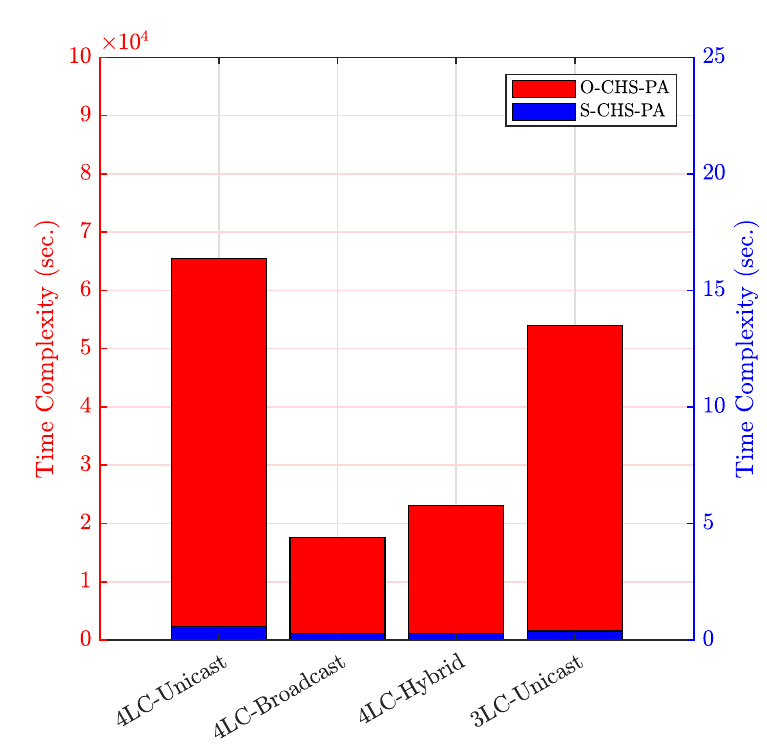}}
\caption{Time complexity of \gls{opa} vs. \gls{gpa} for \gls{udmnoma} with \gls{sic} error = $1\%$.}
    \label{fig:complexity_scenarios}
\end{figure}

 Here, we compare the complexity of \gls{o-chs-pa} against the proposed \gls{s-chs-pa} using the time elapsed feature in MATLAB \cite{Hamad2022-TVT} on a machine equipped with a 3.0 GHz 8-core processor and 16 GB RAM. Figure \ref{fig:complexity_scenarios} shows the convergence time for the optimal and the proposed sub-optimal schemes for the four scenarios using \gls{udmnoma} with 1\% \gls{sic} error. The results show that \gls{o-chs-pa} takes orders of magnitude more time to converge compared to \gls{s-chs-pa}. While \gls{s-chs-pa} converges in under 0.7s for all scenarios, the optimal scheme takes many hours to complete. Notably, the 4LC-Unicast is the most computationally expensive scenario as it includes more variables than the other schemes. Remarkably, the elapsed time measured in MATLAB is affected by running programs and operating system overheads. Hence, a dedicated vehicle computer may achieve near-real-time performance for the S-CHS-PA scheme. Finally, note that for all simulated scenarios, GPA converged in less than 70 iterations, which is below $N_{itr}^{max}$.

\section{Broader Implications and Discussions}\label{sec:Discussion}

\subsection{Comparison of the Proposed NOMA Schemes}\label{sec:comparison}
 The simulation results show that each proposed \gls{m2m} \gls{noma} scheme is superior in sum rate for a given scenario of message type and SIC error. Specifically, \gls{dmnoma} may outperform \gls{umnoma} when there are more receiving LCs than transmitters and vice versa. Conversely, \gls{umnoma} may be more advantageous than \gls{dmnoma} in scenarios including more broadcast/multicast messages than unicast messages. Meanwhile, \gls{udmnoma} excels in all message scenarios unless the \gls{sic} error is too high. While \gls{udmnoma} introduces higher delay in the \gls{sic} process due to the increased number of interference terms,  it requires a single time slot to exchange the \gls{lc} messages. Additionally, the high data rate of \gls{udmnoma} reduces packet accumulation in transmit buffer and hence minimizes packet delay. Although \gls{dmnoma} incurs the same \gls{sic} overhead as \gls{umnoma}, it demands less strict time synchronization among the \glspl{lc}, as only one \gls{ch} transmits at a time.

\subsection{Practical Challenges of the Proposed Schemes}\label{sec:practicality}
The proposed solution entails a number of practical challenges discussed as follows:
\subsubsection{Channel Estimation} \label{sec:practical_csi}
 As the \gls{nch} receivers require complete \gls{csi} to be able to decode the messages, a channel estimation period managed by the \gls{sch} may commence wherein each selected \gls{ch} transmits channel estimation pilots in a \gls{tdma} fashion. Assuming reciprocal channels, only $N_{LC}$ time slots are needed to estimate the channels between each pair of LCs. Note that imperfect \gls{csi} leads to imperfect \gls{sic}. However, the results show that gains over \gls{oma} can still be achieved with some degree of \gls{sic} error.  
\subsubsection{SIC Overhead} \label{sec:practical_sic}
Another practical challenge is the potentially extended \gls{sic} delay and processing overhead due to the large number of \gls{sic} terms in \gls{udmnoma} that reaches $4$ and $9$ terms for $N_{LC}=3$ and $N_{LC}=4$, respectively. Hence, we suggest that \gls{sc} size should be limited depending on the processing capabilities of nodes and latency constraints by dividing the network into many adjacent \glspl{sc}. Despite the delays and overheads discussed above, the \gls{udmnoma} scheme reduces the number of time slots by a factor of $6$ and $12$ for $N_{LC}=3$ and $N_{LC}=4$, respectively. 
\subsubsection{Energy Consumption} \label{sec:practical_energy}
While performing \gls{sic} and solving the \gls{chs-pa} problem require significant processing power, we believe that the computational energy consumed is negligible compared to the energy savings achieved by the proposed vehicular network schemes. This is because a car engine typically consumes power in the range of tens of kilowatts, while an advanced in-car computer, such as those used in Tesla cars, consumes only a few hundred watts. Considering that platooning can reduce fuel consumption by approximately 20\% , the energy savings from reduced fuel use outweigh the power consumed by a high-performance car computer.

Despite the above challenges, NOMA brings many advantages besides the gain in the sum rate. It eliminates time synchronization overheads, as all users can transmit at any time, minimizing latency. Also, using the same frequency band for all transmissions means that channel reciprocity can simplify channel estimation. Finally, NOMA enhances user-fairness and improve spectral efficiency \cite{noma_myths}.

\subsection{Special Network Topologies}\label{sec:topologies}
Finally, we discuss two special network scenarios. The first is when the network consists of only two \glspl{lc}, where \gls{noma} will not be utilized. Yet, this scenario benefits from the full-duplex transmission enabled by the proposed model. Secondly, consider a network with an \gls{lc} containing a single vehicle. As the single-vehicle \gls{lc} cannot transmit and receive simultaneously, the \gls{lc} may transmit a message in a time slot and then receive the other \gls{lc} messages in the following time slot. Note that the aforementioned scenarios generally occur in a low-density network in which the application of our proposed NOMA schemes will not be critical.

\section{Conclusions} \label{sec:Conclusions}
This paper presents  novel NOMA-based techniques for achieving \gls{urllc} and high throughput in cluster-based \glspl{vanet} and C-V2X side link mode. The proposed \gls{udmnoma} scheme allows clusters to exchange unicast, multicast, and broadcast messages in the same \gls{rb}, thus enhancing data rates while minimizing latency. Further, a low-complexity near-optimal greedy algorithm-based solution is designed to solve the sum-rate maximizing \gls{chs-pa} problem. A proposed SC formation protocol maintains the SC size relatively small and a function of the processing capabilities of the participating nodes and the latency constraints. The presented results demonstrate that \gls{udmnoma} outperforms \gls{oma} by at least 50\% despite a moderate \gls{sic} error. While \gls{udmnoma} is generally superior, \gls{umnoma} and \gls{dmnoma} may occasionally achieve higher sum rates. Therefore, the \gls{noma} scheme should be selected by the \gls{sch} based on the specific system parameters, application scenario, and \gls{qos} requirements by means of an algorithm that we aim to develop in future work. Additionally, we aim to develop the \gls{scfp} in a network simulator, such as OMNET++, and evaluate its performance in a wide range of real-world vehicular scenarios. Finally, a more efficient \gls{chs-pa} algorithm based on deep learning or game theory may be developed.                

\bibliographystyle{IEEEtran}

\end{document}